\documentclass[11pt, a4paper]{article}

\newif\ifpublic\publictrue
\newif\iffancy\fancytrue

\usepackage{jheppub}
\usepackage{amsmath,amstext,amssymb,amscd,mathtools}
\usepackage{graphicx}
\usepackage[usenames,dvipsnames]{xcolor}
\usepackage{bbm}
\usepackage{enumitem}
\usepackage{circledsteps}
\usepackage{adjustbox}
\usepackage{tikz}
\usetikzlibrary{arrows, positioning}
\title{Exploring Reggeon bound states in strongly-coupled $\mathcal{N}=4$ super Yang-Mills}
\author[a, b]{Theresa Abl}
\emailAdd{theresaabl@gmx.at}
\author[b]{and Martin Sprenger}
\affiliation[a]{Department of Mathematical Sciences, Durham University,\\Durham DH1 3LE, United Kingdom}
\affiliation[b]{Institut f\"ur Theoretische Physik, Eidgen\"ossische Technische Hochschule Z\"urich,\\Wolfgang-Pauli-Strasse 27, 8093 Z\"urich, Switzerland}
\emailAdd{sprenger-m@gmx.de}
\abstract{The multi-Regge limit of scattering amplitudes in strongly-coupled $\mathcal{N}=4$ super Yang-Mills is described by the large mass limit of a set of thermodynamic Bethe ansatz (TBA) equations. A non-trivial remainder function arises in this setup in certain kinematical regions due to excitations of the TBA equations which appear during the analytic continuation into these kinematical regions. So far, these analytic continuations were carried out on a case-by-case basis for the six- and seven-gluon remainder function. In this note, we show that the set of possible excitations appearing in any analytic continuation in the multi-Regge limit for any number of particles is rather constrained. In particular, we show that the BFKL eigenvalue of any possible Reggeon bound state is a multiple of the two-Reggeon BFKL eigenvalue appearing in the six-gluon case.}
\begin{document}

\newcommand{\Y}[2]{\mathrm{Y}_{#1}^{\left[#2\right]}}
\newcommand{\Yp}[2]{\mathrm{Y}_{#1}^{\prime\left[#2\right]}}
\newcommand{\U}[2]{\frac{\Y{2}{#1}{#2}}{1+\Y{2}{#1}{#2}}}
\newcommand{\Up}[2]{\frac{\Yp{2}{#1}{#2}}{1+\Yp{2}{#1}{#2}}}
\newcommand{\cratio}[4]{\frac{x_{#1,#2}^2 x_{#3,#4}^2}{x_{#3,#2}^2 x_{#1,#4}^2}}
\newcommand{\Yf}{\mathrm{Y}}
\newcommand{\Yt}{\widetilde{\Yf}}

\setcounter{tocdepth}{2}
\maketitle
\section{Introduction}
\label{sec:intro}
In the context of the planar limit of $\mathcal{N}=4$ super Yang-Mills (SYM) theory, the development of new techniques has led to remarkable progress being made for the calculation of scattering amplitudes, pushing the known results for the six- and seven-point gluon scattering amplitudes up to seven and four loops, respectively \cite{Bern:2005iz, DelDuca:2010zg, Dixon:2011pw, CaronHuot:2011kk, Dixon:2011nj, Dixon:2013eka, Dixon:2014voa, Dixon:2014iba, Drummond:2014ffa, Dixon:2015iva, Caron-Huot:2016owq, Dixon:2016nkn, Drummond:2018caf, Caron-Huot:2019vjl, Caron-Huot:2019bsq, Dixon:2020cnr}.
Much less is still known, however, about scattering amplitudes with more external gluons, at least in general kinematics, where only the two-loop symbol is known for any number of gluons \cite{CaronHuot:2011ky, Zhang:2019vnm, He:2020vob}, and the remainder function of the eight- and nine-gluon amplitude was determined up to two loops in \cite{Golden:2021ggj}.
In this situation, considering the multi-Regge limit can provide valuable insights on the path to the amplitude in full kinematics due to (at least) two of its properties which were historically derived starting with the seminal BFKL papers \cite{Lipatov:1976zz, Fadin:1975cb, Kuraev:1976ge, Balitsky:1978ic}:
\begin{itemize}
	\item Firstly, the multi-Regge limit naturally re-organizes the perturbative expansion of the scattering amplitude from loop levels to an expansion in logarithmic accuracy, where each order corresponds to a resummation of certain large logarithms appearing in this limit.
This expansion starts with the \textit{leading logarithmic approximation} (LLA) and continues with the \textit{(Next-to-)}$^k$-LLA (N$^k$LLA), with each logarithmic order containing information from all loop orders, thus providing valuable information on the all-loop structure of the amplitude, albeit in special kinematics.
The multi-Regge limit is naturally formulated in terms of effective particles, so-called Reggeons.
Scattering amplitudes can then be described by quantities related to those Reggeons, namely BFKL eigenvalues describing the propagation of bound states of two or more Reggeons and impact factors describing the transition of one Reggeon bound state to another.
	\item Secondly, the BFKL eigenvalues and impact factors are universal quantities.
This means, for example, that the BFKL eigenvalue describing a two-Reggeon bound state is always the same, independent of whether it appears in a six-, seven- or $n$-point amplitude.
Thus, results obtained for a lower-point amplitude can be re-used in higher-point amplitudes.
Furthermore, the appearance of BFKL eigenvalues and impact factors is closely linked to the concept of so-called Mandelstam regions, which are kinematical regions reached by analytic continuations of the scattering amplitude in the kinematic variables.
The theory of the multi-Regge limit predicts which of these quantities appear in which Mandelstam region, thus, providing important data for a potential ansatz in general kinematics.
\end{itemize}
The ultimate goal of the study of the multi-Regge limit in planar $\mathcal{N}=4$ SYM therefore is to determine all possible BFKL eigenvalues and impact factors exactly.
Our paper presents a small step towards this goal by determining all possible BFKL eigenvalues at strong coupling.\par
In this regard, the BFKL eigenvalue of the two-Reggeon bound state and the impact factor describing the transition from one Reggeon to a bound state of two Reggeons were determined perturbatively up to N$^2$LLA \cite{Bartels:2008ce, Bartels:2008sc, Lipatov:2010ad, Fadin:2011we, Dixon:2012yy, Dixon:2013eka, Dixon:2014voa} and at strong coupling \cite{Bartels:2010ej, Bartels:2013dja}, before finite coupling expressions were derived in \cite{Basso:2014pla}.
These ingredients are sufficient to describe the six-gluon amplitude.
However, starting from seven gluons, an additional quantity, namely the central emission vertex, is needed to describe the amplitude in all kinematic regions.
This quantity describes the emission of a physical gluon from the two-Reggeon bound state.
The central emission vertex was determined perturbatively up to NLLA \cite{Bartels:2011ge, Bartels:2013jna, Bartels:2014jya, DelDuca:2018hrv} and its contribution at strong coupling was analyzed in \cite{Bartels:2014ppa, Bartels:2014mka, Sprenger:2016jtx}, before a conjecture for an exact expression at finite coupling was presented in \cite{DelDuca:2019tur}.
As the number of gluons increases, further ingredients will be required for the calculation of a $n$-gluon amplitude in the multi-Regge limit.
This includes, in particular, the BFKL eigenvalue of Reggeon bound states consisting of more than two Reggeons.
From Regge theory, the BFKL eigenvalue of a $k$-Reggeon bound state is expected to appear starting from the $2k+2$-gluon amplitude \cite{Lipatov:2009nt, Bartels:2011nz}.
Understanding these Reggeon bound states is hence crucial in pushing the exact results known in the multi-Regge limit for the six- and seven-point amplitudes to higher-point amplitudes.
However, compared to the quantities appearing in the six- and seven-point amplitude, very little is known regarding the BFKL eigenvalues of bound states of more than two Reggeons.
So far, the contribution of a three-Reggeon bound state was first observed in the context of the two-loop eight-point remainder function \cite{DelDuca:2018raq}, and its contribution to the eight-point remainder function from the perspective of Regge theory is systematically explored in \cite{Bartels:2020twc}.\par
A rather simple picture for the Reggeon bound states is suggested by the Wilson loop OPE \cite{Alday:2010ku, Gaiotto:2010fk, Gaiotto:2011dt, Sever:2011da, Basso:2013vsa, Basso:2015uxa}, which is based on an expansion of the remainder function around the collinear limit.
This expansion is described by an integrable flux-tube spanned by a Wilson loop, on which excitations propagate.
The properties of these excitations, such as their dispersion relations and their scattering matrices are known at finite coupling \cite{Basso:2010in, Basso:2013aha, Belitsky:2014rba, Basso:2014koa, Basso:2014nra, Belitsky:2014sla, Belitsky:2014lta, Basso:2014hfa, Belitsky:2015efa, Basso:2015rta}, which in principle allows an evaluation of the remainder function to arbitrary precision, although in practice the resummation of the excitations is difficult and was so far only carried out in individual cases (see, for example, \cite{Drummond:2015jea, Fioravanti:2015dma, Bonini:2015lfr, Cordova:2016woh, Lam:2016rel, Basso:2020xts}).
To connect the collinear limit and the multi-Regge limit, an analytic continuation of the Wilson loop OPE from the collinear to the multi-Regge regime for the six-point amplitude is used in \cite{Basso:2014pla} to extract exact expressions for the two-Reggeon BFKL eigenvalue and the impact factor.
This method was extended in \cite{DelDuca:2019tur} to propose an exact expression for the central emission vertex.
In this approach, the BFKL eigenvalue corresponds to an analytic continuation of the energy of certain excitations.
Since the system described by these excitations is integrable, the energy of several of these excitations corresponds to the sum of the energies of the individual excitations.
Hence, there seems to be no room for additional, fundamentally new structures appearing at higher-point amplitudes.
Assuming that the multi-Regge limit of higher-point amplitudes can also be described by a continuation from the collinear limit, this picture suggests that the BFKL eigenvalues of bound states of more than two Reggeons are, in fact, not new and independent quantities, but should be related to the already known two-Reggeon bound state.
This is in line with results from Regge theory, where, to leading order, the BFKL eigenvalue is identified with the energy of a spin chain model, for which the number of sites is determined by the number of Reggeons in the bound state \cite{Lipatov:2009nt, Bartels:2011nz}.
However, from the perspective of Regge theory, it is not at all obvious whether such a simple picture prevails beyond leading order, as well (see, for example, \cite{Bartels:2012sw}).\par
We approach this question by calculating the possible BFKL eigenvalues for any number of gluons at strong coupling, where the remainder function is determined by a set of thermodynamic Bethe ansatz (TBA) equations.
The analytic continuation of these TBA equations in the multi-Regge limit connects each Mandelstam region with a set of coordinate Bethe ansatz equations, which then determines the remainder function.
In previous papers, the calculation of amplitudes in the multi-Regge limit at strong coupling was based on a case-by-case analysis, in which the analytic continuation of the TBA equations to different kinematic regions was carried out numerically \cite{Bartels:2010ej, Bartels:2013dja, Bartels:2014ppa, Bartels:2014mka, Sprenger:2016jtx}.
This approach becomes more and more difficult as the number of gluons increases, as the number of kinematic variables which need to be analytically continued grows and since the choice of the correct paths of analytic continuation becomes more involved (see, for example, \cite{Bartels:2014mka, Bargheer:2015djt, Bargheer:2019lic}).
Therefore, we take a different route in this paper.
Rather than performing the analytic continuations explicitly, we show that the multi-Regge limit imposes strong constraints on the allowed Bethe ansatz equations.
We then determine the remainder function for the Bethe ansatz equations allowed by the multi-Regge limit and show that the resulting BFKL eigenvalues are always multiples of the two-Reggeon BFKL eigenvalue.
Furthermore, we show that this result also holds for a set of kinematically subleading terms, which were analyzed in the six-point case in \cite{Sprenger:2016jtx} and which yield information beyond the strong coupling saddle point.
Thus, our result provides evidence that the BFKL eigenvalues of bound states of three or more Reggeons are indeed simple functions of the two-Reggeon BFKL eigenvalue, at least at strong coupling.\par
This paper is organized as follows.
In section \ref{sec:review} we review the calculation of scattering amplitudes in planar $\mathcal{N}=4$ SYM at strong coupling and explain that the remainder function in the multi-Regge is determined by certain singularities of the associated TBA equations.
We then discuss the behavior of these singularities during an arbitrary analytic continuation and the constraints the multi-Regge limit imposes on the relevant singularities in section \ref{sec:analysis_bae}, before calculating the remainder function for the most general Bethe ansatz allowed by the multi-Regge limit in section \ref{sec:rem_fct}.
We apply our results to the specific case of the nine-point amplitude in section \ref{sec:three_reggeon_bs}, in which we also analyze the structure of kinematically subleading contributions to the remainder function.
Lastly, we summarize our results and discuss open questions in section \ref{sec:conclusions}.
Technical results are collected in several appendices.

\section{Review: The multi-Regge limit in strongly-coupled $\mathcal{N}=4$ SYM}
\label{sec:review}
\subsection{Scattering amplitudes at strong coupling}
\label{sec:review_tba}
Let us begin by briefly reviewing the calculation of scattering amplitudes in strongly-coupled planar $\mathcal{N}=4$ SYM as derived in \cite{Alday:2007hr, Alday:2009yn, Alday:2009dv, Alday:2010vh}.
At strong coupling, the color-ordered $n$-gluon amplitude is given by
\begin{equation}
	\mathcal{A}_{n}\sim e^{-\frac{\sqrt{\lambda}}{2\pi} A_{\mathrm{BDS}}(x_i)+R_n(u_{a,s})},
	\label{eq:def_amplitude}
\end{equation}
to leading order in $\sqrt{\lambda}$, where $\lambda$ is the 't Hooft coupling constant.
In equation (\ref{eq:def_amplitude}), $A_{\mathrm{BDS}}$ is the strong coupling limit of the BDS ansatz \cite{Bern:2005iz}, which contains the IR-divergences of the amplitude and $R_n$ is the so-called remainder function.\footnote{Note that at strong coupling, the dependence of the remainder function on the helicity of the gluons is subleading in $\sqrt{\lambda}$. Therefore, $R_n$ in eq.\ (\ref{eq:def_amplitude}) is the same for all N$^{k}$MHV configurations.}
To describe the kinematic dependence in eq.\ (\ref{eq:def_amplitude}), we introduce the dual variables $x_i$ by $p_i=:x_{i-1}-x_{i}$, with $x_{i+n}\equiv x_i$, where $p_i$ are the gluon momenta.
While the BDS ansatz in eq.\ (\ref{eq:def_amplitude}) depends explicitly on the $x_i$, the remainder function only depends on $3n-15$ cross ratios $u_{a,s}$ due to dual conformal symmetry \cite{Drummond:2008vq}, for which we choose the basis
\begin{equation}
	u_{1,s}:=\cratio{s+1}{s+5}{s+2}{s+4},\quad u_{2,s}:=\cratio{1}{s+2}{n}{s+3},\quad u_{3,s}:=\cratio{2}{s+3}{1}{s+4},
	\label{eq:def_basis_crs}
\end{equation}
where $x_{i,j}:=x_i-x_j$ and $s=1,\dots, n-5$.
The dependence of the remainder function on the cross ratios is described by three terms,\footnote{Note that for the case $n=4k$, an additional contribution $A_{\mathrm{extra}}$ to the remainder function exists \cite{Alday:2009yn, Yang:2010az, Yang:2010as}, which we ignore in the following. We show in section \ref{sec:rem_fct} that this does not limit the validity of our results.}
\begin{equation}
	R_n:=-\frac{\sqrt{\lambda}}{2\pi}\left(\Delta+A_{\mathrm{per}}+A_{\mathrm{free}}\right),
	\label{eq:def_rem_fct}
\end{equation}
which we discuss in turn.
The simplest term, $\Delta=\Delta(u_{a,s})$, is a transcendentality-two function of the cross ratios.
Explicit expressions are derived, for example, in \cite{Yang:2010as}.
To explain the other two terms in eq.\ (\ref{eq:def_rem_fct}), we introduce the auxiliary parameters $m_s=:|m_s|e^{i\varphi_s}$ and $C_s$, which are connected with the cross ratios (\ref{eq:def_basis_crs}) as described below.
We furthermore introduce $3n-15$ functions $\widetilde{\Yf}_{a,s}(\theta)$, which depend on the auxiliary parameters and a complex parameter $\theta$.
These $\widetilde{\Yf}$-functions satisfy the non-linear integral equations
\begin{equation}
	\log\widetilde{\Yf}_{a,s}(\theta)=-|m_{a,s}|\cosh \theta+C_{a,s}+\sum\limits_{a',s'}\int\limits_{\mathbb{R}} d\theta'\,\mathcal{K}_{s,s'}^{a,a'}(\theta-\theta'+i\varphi_s-i\varphi_{s'})\log\left(1+\widetilde{\Yf}_{a',s'}(\theta')\right),
	\label{eq:def_tba}
\end{equation}
where $a, a'\in\{1,2,3\}$, $s'\in\{s-1, s, s+1\}$ and $s=1, \dots, n-5$ as before.
These equations resemble thermodynamic Bethe ansatz equations and are called the $\Yf$-system or the TBA equations in the following.
In eq.\ (\ref{eq:def_tba}), we have introduced the quantities
\begin{equation}
	|m_{a,s}|:=\sqrt{2}^{\,\delta_{a,2}}|m_s|,\quad C_{a,s}:=\left(\delta_{a,3}-\delta_{a,1}\right)C_s,
	\label{eq:def_params_tba}
\end{equation}
and the integration kernels $\mathcal{K}_{s,s'}^{a,a'}$, which are hyperbolic functions spelled out explicitly in appendix \ref{app:kernels}.
Eq.\ (\ref{eq:def_tba}) defines the $\widetilde{\Yf}_{a,s}$-functions for small imaginary parts of $\theta+i\varphi_s-i\varphi_{s'}$.
Whenever $\mathrm{Im}\left(\theta+i\varphi_s-i\varphi_{s'}\right)=k\cdot \frac{\pi}{4}$ with $k\in\mathbb{Z}$ holds, singularities of the integration kernels lie on the integration contour and the corresponding residues need to be picked up, which introduces additional terms in eq.\ (\ref{eq:def_tba}).
As an alternative to picking up the residues of the singularities appearing for large imaginary parts of $\theta$ explicitly, one can use the recursion relations
\begin{equation}
	\widetilde{\Yf}_{a,s}(\theta)=\frac{\left(1+\widetilde{\Yf}_{a,s+1}^{[\pm 1]}(\theta+i\varphi_s-i\varphi_{s+1})\right) \left(1+\widetilde{\Yf}_{4-a,s-1}^{[\pm 1]}(\theta+i\varphi_s-i\varphi_{s-1})\right)}{\widetilde{\Yf}_{4-a,s}^{[\pm 2]}(\theta)\left(1+\frac{1}{\widetilde{\Yf}_{a+1,s}^{[\pm 1]}(\theta)}\right)\left(1+\frac{1}{\widetilde{\Yf}_{a-1,s}^{[\pm 1]}(\theta)}\right)},	
	\label{eq:rec_rels}
\end{equation}
where $\widetilde{\Yf}_{a,s}^{[\pm k]}(\theta):=\widetilde{\Yf}_{a,s}\left(\theta\pm ik\frac{\pi}{4}\right)$, subject to the boundary conditions 
\begin{equation}
	\widetilde{\Yf}_{0, s}(\theta)=\widetilde{\Yf}_{4,s}(\theta)=\infty,\quad\mathrm{and}\quad\widetilde{\Yf}_{a,0}(\theta)=\widetilde{\Yf}_{a, n-4}(\theta)=0
	\label{eq:bdy_cond}
\end{equation}
to relate the $\widetilde{\Yf}_{a,s}$-functions for any value of $\theta$ with those closer to the real axis.\par
To calculate the $A_{\mathrm{free}}$-contribution to the remainder function (\ref{eq:def_rem_fct}) for given parameters $m_s$ and $C_s$, one first needs to solve eqs.\ (\ref{eq:def_tba}) for the $\widetilde{\Yf}_{a,s}$-functions on the real line, which then are used to determine $A_{\mathrm{free}}$,
\begin{equation}
	A_{\mathrm{free}}=\sum\limits_{a,s}\frac{|m_{a,s}|}{2\pi}\int\limits_{\mathbb{R}} d\theta \cosh\theta\,\log\left(1+\widetilde{\Yf}_{a,s}(\theta)\right).
	\label{eq:a_free}
\end{equation}
Lastly, $A_{\mathrm{per}}=A_{\mathrm{per}}(m_s)$ is simply a polynomial in the auxiliary parameters $m_s$, which we spell out for particular amplitudes in later sections.\par
This procedure allows to us to calculate the remainder function for given values of the auxiliary parameters $m_s$ and $C_s$.
However, we still need to connect the auxiliary parameters with the cross ratios which are used to describe the kinematics.
The relation between the two sets of parameters is derived in \cite{Alday:2010vh} and reads
\begin{align}
	u_{1,s}=\frac{\Yf_{2,1}^{[2s+7]}}{1+\Yf_{2,1}^{[2s+7]}},\quad u_{2,s}=\frac{\Yf_{2,s}^{[s+4]}}{1+\Yf_{2,s}^{[s+4]}},\quad u_{3,s}=\frac{\Yf_{2,s}^{[s+6]}}{1+\Yf_{2,s}^{[s+6]}},
	\label{eq:rel_aux_crs}
\end{align}
where $\Yf_{a,s}^{[k]}:=\widetilde{\Yf}_{a,s}(ik\frac{\pi}{4}-i\varphi_s)$, for the chosen basis (\ref{eq:def_basis_crs}).
To calculate the remainder function in practice, we would thus need to specify the values of the cross ratios we are interested in and then try to find those values of the auxiliary parameters that reproduce the behavior of the cross ratios via eq.\ (\ref{eq:rel_aux_crs}).
\subsection{The multi-Regge limit of the $\Yf$-system}
\label{sec:mrl_tba}
\begin{figure}[t]
	\centering
	\begin{minipage}[b]{.45\textwidth}
		\centering
		\includegraphics{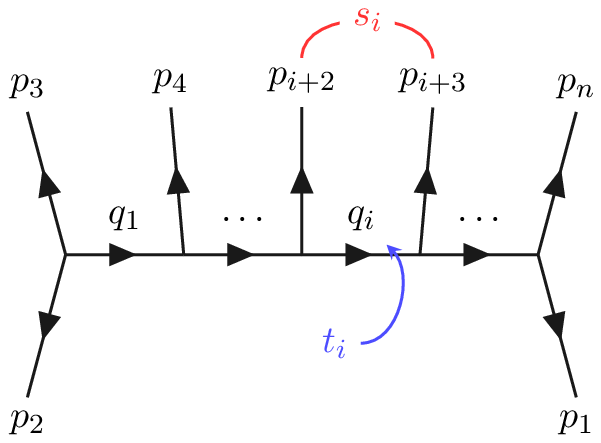}
	\end{minipage}\hfill
	\begin{minipage}[b]{.45\textwidth}
		\centering
		\includegraphics{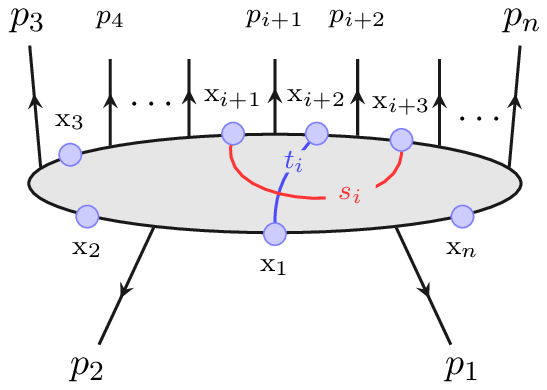}
	\end{minipage}
	\caption{Graphical representation of the Mandelstam variables $s_i$ (\ref{eq:def_mandelstam_s}) and $t_i$ (\ref{eq:def_mandelstam_t}) in standard momenta (left) and dual momenta (right) using the all-outgoing convention.}
	\label{fig:kinematics}
\end{figure}

So far, our discussion of the $\Yf$-system holds in general kinematics, i.e.\ for general values of the cross ratios and the auxiliary parameters.
However, in this setting the TBA equations (\ref{eq:def_tba}) are difficult to solve and do not have an analytic solution.
Therefore, we now consider a special kinematical configuration, namely the multi-Regge limit (MRL), which is characterized by the limit in which the rapidities of the outgoing particles are strongly ordered.
In terms of Mandelstam variables, this translates into the limit in which the $s$-like Mandelstam variables 
\begin{equation}
	s_i:=(p_{i+2}+p_{i+3})^2=x_{i+1,i+3}^2,\quad i=1,\dots, n-3,
	\label{eq:def_mandelstam_s}
\end{equation}
of all channels become large while the corresponding $t$-like momentum transfers 
\begin{equation}
	t_i:=(p_{2}+\dots+p_{i+2})^2=x_{1,i+2}^2
	\label{eq:def_mandelstam_t}
\end{equation}
remain finite (see, for example, \cite{Bartels:2012gq} for a derivation).
Furthermore, we restrict our attention to $2\rightarrow n-2$ scattering, so that $p_1$, $p_2$, and $p_3$ and $p_n$ describe the initial and final momenta of the scattering particles, respectively, while $p_4,\dots,p_{n-1}$ describe the gluons produced in the scattering process (cf.\ figure \ref{fig:kinematics}).
In terms of the cross ratios, the multi-Regge limit corresponds to the limit
\begin{equation}
	u_{1,s}\rightarrow 1,\quad u_{2,s}\rightarrow 0,\quad u_{3,s}\rightarrow 0,
	\label{eq:def_mrl}
\end{equation}
where the limit is taken in such a way that the so-called reduced cross ratios $\tilde{u}_{2/3,s}:=\frac{u_{2/3, s}}{1-u_{1,s}}$ attain finite values in the limit. Inspired by eq.~\eqref{eq:def_mrl} the cross ratios $u_{1,s}$ are called \textit{large} cross ratios, while $u_{2/3,s}$ are called \textit{small} cross ratios. 
The limit is taken independently for different values of $s$, which corresponds to the sub-energies $s_i$ of the different channels becoming large independently.\par
In \cite{Bartels:2010ej, Bartels:2012gq} it is shown that in terms of the auxiliary parameters of the $\Yf$-system, the multi-Regge limit corresponds to the limit
\begin{equation}
	|m_s|\rightarrow\infty,\quad \varphi_s\rightarrow (1-s)\frac{\pi}{4},\quad C_s\rightarrow \mathrm{const.}\, ,
	\label{eq:mrl_aux}
\end{equation}
where the $C_s$ are purely imaginary.
This is a special limit from the perspective of the TBA, as all integrals are exponentially suppressed due to the parameters $|m_{a,s}|$ becoming large, and the equations (\ref{eq:def_tba}) reduce to algebraic equations,
\begin{equation}
	\log\widetilde{\Yf}_{a,s}(\theta)\cong -|m_{a,s}|\cosh\theta+C_{a,s},
	\label{eq:yfct_fund_strip}
\end{equation}
where `$\cong$' denotes equality in the multi-Regge limit, and are therefore much easier to work with than in general kinematics. Note, however, that this simple form only holds close to the real axis.
For large imaginary parts, one still has to pick up the residue contributions of the integration kernels before dropping the integrals, or use the recursion relations (\ref{eq:rec_rels}).
An analysis carried out in \cite{Bartels:2012gq} shows that due to the values of the parameters $\varphi_s$ in the multi-Regge limit (\ref{eq:mrl_aux}), the simple form of the TBA equations (\ref{eq:yfct_fund_strip}) strictly speaking only holds for real $\theta$.
However, as shown in that reference, the residue contributions of the integration kernels are negligible in the region $-\frac{\pi}{4}\leq \mathrm{Im}\,\theta\leq\frac{\pi}{4}$, so that the simple form of the TBA equations actually holds within this region and we refer to this region as the \textit{fundamental strip} in the following.
This result is crucial for our analysis of the possible Bethe ansatz equations in section \ref{sec:constraints}.
\par
For the values of the parameters $\varphi_s$ in the MRL as shown in eq.\ (\ref{eq:mrl_aux}), the recursion relations (\ref{eq:rec_rels}) simplify.
For example, the recursion relation in the direction of decreasing $\mathrm{Im}\,\theta$ reads 
\begin{equation}
	\widetilde{\Yf}_{a,s}(\theta)=\frac{\left(1+\widetilde{\Yf}_{a,s+1}(\theta)\right) \left(1+\widetilde{\Yf}_{4-a,s-1}\left(\theta-i\frac{\pi}{2}\right)\right)}{\widetilde{\Yf}_{4-a,s}\left(\theta-i\frac{\pi}{2}\right)\left(1+\frac{1}{\widetilde{\Yf}_{a+1,s}(\theta-i\frac{\pi}{4})}\right)\left(1+\frac{1}{\widetilde{\Yf}_{a-1,s}(\theta-i\frac{\pi}{4})}\right)},	
	\label{eq:rec_rels_mrl}
\end{equation}
and a similar recursion relation holds in the direction of increasing $\mathrm{Im}\,\theta$.
Furthermore, in the MRL the cross ratios (\ref{eq:def_basis_crs}) correspond to the following $\widetilde{\Yf}$-functions:
\begin{align}
	u_{1,s}=\frac{\widetilde{\Yf}_{2,1}^{[2s+7]}}{1+\widetilde{\Yf}_{2,1}^{[2s+7]}},\quad u_{2,s}=\frac{\widetilde{\Yf}_{2,n-4-s}^{[-1]}}{1+\widetilde{\Yf}_{2,n-4-s}^{[-1]}},\quad u_{3,s}=\frac{\widetilde{\Yf}_{2,n-4-s}^{[1]}}{1+\widetilde{\Yf}_{2,n-4-s}^{[1]}},
	\label{eq:rel_aux_crs_mrl}
\end{align}
where we have used the symmetry $\Yf_{2,s}^{[k]}=\Yf_{2,n-4-s}^{[k\pm n]}$ to obtain these equations from eq.\ (\ref{eq:rel_aux_crs}).\footnote{This symmetry is a consequence of a symmetry of the Hitchin system underlying the $\Yf$-system, see \cite{Alday:2010vh}.}
In particular, the small cross ratios $u_{2/3,s}$ are located at fixed values of $\theta$, which simplifies their evaluation in the multi-Regge limit.
For this purpose, we introduce the parameters
\begin{equation}
\varepsilon_s\coloneqq e^{-|m_s| \cos\left((s-1)\frac{\pi}{4}+\varphi_s\right)},\quad w_s\coloneqq e^{|m_s| \sin\left((s-1)\frac{\pi}{4}+\varphi_s\right)},
\label{eq:def_mrl_param}
\end{equation}
which show the following behavior in the multi-Regge limit:
\begin{equation}
\varepsilon_s \rightarrow 0,\qquad w_s\rightarrow \text{const.}
\end{equation}
In terms of these parameters, the cross ratios behave as
\begin{align}
	u_{1,s}&=1-\varepsilon_{n-4-s}\left(w_{n-4-s}+\frac{1}{w_{n-4-s}}+2\cosh C_{n-4-s}\right),\nonumber\\
	u_{2,s}&=\varepsilon_{n-4-s}\cdot w_{n-4-s},\nonumber\\
	u_{3,s}&=\frac{\varepsilon_{n-4-s}}{w_{n-4-s}}
	\label{eq:crs_MRL}
\end{align}
in the multi-Regge limit with corrections of order $\mathcal{O}(\varepsilon^2)$, see \cite{Bartels:2012gq} for details.
This is precisely the behavior required by the multi-Regge limit (\ref{eq:def_mrl}).\par
Due to the simple form of the $\widetilde{\Yf}_{a,s}$-functions in the fundamental strip (\ref{eq:yfct_fund_strip}), the integrals in the $A_{\mathrm{free}}$-contribution to the remainder function (\ref{eq:a_free}) are also negligible.
In fact, the whole remainder function is trivial in the limit (\ref{eq:mrl_aux}),
\begin{equation}
	A_{\mathrm{free}}+A_{\mathrm{per}}+\Delta \rightarrow \mathrm{const.}, 
	\label{eq:rem_fct_mrl}
\end{equation}
see \cite{Bartels:2010ej, Bartels:2012gq, Bartels:2014mka} for details.
This, however, is not the end of the story as taking the limit (\ref{eq:mrl_aux}) of the remainder function (\ref{eq:def_rem_fct}) corresponds to the multi-Regge limit in the Euclidean region only, in which the remainder function is known to vanish at weak coupling, too \cite{Lipatov:2009nt, Bartels:2011nz}.\par
As mentioned in section \ref{sec:intro}, we need to consider other kinematical regions to unravel the full structure of the remainder function in the multi-Regge limit.
The transition to other kinematical regions (so-called Mandelstam regions) is described by an analytic continuation of the remainder function in the cross ratios (\ref{eq:def_basis_crs}).
In this paper, we are deliberately vague regarding the precise form of the paths of analytic continuation.
Indeed, our goal is to show that we can determine the possible BFKL eigenvalues without specifying the path of continuation explicitly.
However, the multi-Regge limit imposes certain constraints on the possible endpoints of the analytic continuations: the paths of analytic continuation are constrained such that the large cross ratios $u_{1,s}$ perform an integer number of rotations around the point $u_{1,s}=0$, while the small cross ratios $u_{2/3, s}$  have winding number $\pm\frac{1}{2}$ and thus may only change their sign at the endpoint of these analytic continuations.
This entails that at the endpoint of any analytic continuation
\begin{equation}
	u_{1,s}'=u_{1,s},\quad u_{2,s}'=\pm u_{2,s},\quad u_{3,s}'=\pm u_{3,s}
	\label{eq:endp_paths_mrl}
\end{equation}
holds for all relevant kinematic regions, where the prime indicates the value of the cross ratios at the endpoint of the continuation. 
The fact that we do not need to specify an explicit path of continuation is a virtue of our approach, since the selection of the correct paths of continuation was a key problem in the calculation of the seven-point amplitude at strong coupling \cite{Bartels:2014mka}.
One problem, for example, is the appearance of dependent cross ratios, which are related to our basis of independent cross ratios (\ref{eq:def_basis_crs}) through conformal Gram relations \cite{Eden:2012tu}, which need to be satisfied throughout the continuation.
This problem becomes worse with increasing number $n$ of gluons, as the number of dependent cross ratios increases, too.
Further aspects regarding the construction of the correct paths of analytic continuation are discussed in \cite{Bartels:2014mka, Bargheer:2015djt, Bargheer:2019lic}.\par
Where required, we denote a specific kinematic region by the signs of the energies of the $n-4$ produced particles, so that, for example, the two-Reggeon bound state contributes to the six-point remainder function in the Mandelstam region $(--)$, and the three-Reggeon bound state is expected to appear in the eight-point remainder function in the region $(-++-)$.
In this paper, we do not consider regions in which the energies of the particles $p_1$, $p_2$, $p_3$ or $p_n$ are analytically continued.
We now continue by exploring the consequences of an analytic continuation of the cross ratios for the $\Yf$-system in the next section.
\subsection{Analytic continuation of the $\Yf$-system}
\label{sec:rev_ac}
An analytic continuation of the cross ratios corresponds to an analytic continuation of the auxiliary parameters $m_s$ and $C_s$, as those encode the kinematics for the $\Yf$-system.
Prescribing a path of analytic continuation for the auxiliary parameters and determining the resulting paths of the cross ratios is simple due to the relations (\ref{eq:rel_aux_crs}).
However, in practice we need to solve the inverse problem, namely prescribing a path of analytic continuation of the cross ratios and finding the corresponding paths of the auxiliary parameters, which is much more difficult as the auxiliary parameters enter the relations (\ref{eq:rel_aux_crs}) only implicitly through the $\widetilde{\Yf}_{2,s}$-functions.
Indeed, determining the correct paths of analytic continuation is a key problem in the explicit calculation of the six- and seven-point remainder function at strong coupling in the multi-Regge limit \cite{Bartels:2010ej, Bartels:2013dja, Bartels:2014ppa, Bartels:2014mka, Sprenger:2016jtx}.
As explained before, we will not specify the paths of analytic continuation for the auxiliary parameters explicitly, as we argue that, as long as the endpoint of the analytic continuation corresponds to a Mandelstam region, the possible BFKL eigenvalues governing the remainder function can be determined without knowing the explicit path chosen for the auxiliary parameters.\par
The analytic continuation of TBA equations was originally studied in the context of other TBAs in \cite{Dorey:1996re, Dorey:1997rb} and is described in the context of the calculation of the remainder function at strong coupling in \cite{Bartels:2010ej, Bartels:2014mka}.
Therefore, we only review the essential pieces and refer the reader to those references for details.
For every $\widetilde{\Yf}_{a,s}$-function, there are special locations $\tilde{\theta}_{a,s}$ at which
\begin{equation}
	\widetilde{\Yf}_{a,s}(\tilde{\theta}_{a,s})=-1
	\label{eq:def_theta_tilde}
\end{equation}
holds.
The location of these points $\tilde{\theta}_{a,s}$, of course, depends on the auxiliary parameters $m_{a,s}$ and $C_{a,s}$.
Hence, as we perform an analytic continuation in the auxiliary parameters, the positions $\tilde{\theta}_{a,s}$ are moving in the complex $\theta$-plane.
The reason these points are special is that they are singularities of the integrand in the $\Yf$-system equations (\ref{eq:def_tba}).
Therefore, if during the analytic continuation one or several of the points $\tilde{\theta}_{a,s}$ cross the integration contour, we have to pick up the corresponding residue contributions.
Parametrizing the solutions of eq.\ (\ref{eq:def_theta_tilde}) which have crossed the integration contour during the analytic continuation as $\tilde{\theta}_{a,s,i}$, where $i=1,\dots,n_{a,s}$ denotes the number of crossing singularities of $\widetilde{\Yf}_{a,s}$, this results in a modified $\Yf$-system
\begin{align}
		\log\widetilde{\Yf}'_{a,s}(\theta)=&-|m_{a,s}|'\cosh\theta+C'_{a,s}+\sum\limits_{a',s'}\int\limits_{\mathbb{R}}d\theta'\mathcal{K}^{a,a'}_{s,s'}(\theta-\theta'+i\varphi'_s-i\varphi'_{s'})\log\left(1+\widetilde{\Yf}'_{a',s'}(\theta')\right)\nonumber\\
		&\quad+\sum\limits_{a',s'}\sum\limits_{i=1}^{n_{a',s'}} \mathrm{sign}(\mathrm{Im}\,\widetilde{\theta}_{a',s',i})\log \mathcal{S}^{a,a'}_{s,s'}\left(\theta-\tilde{\theta}_{a',s',i}+i\varphi_s'-i\varphi_{s'}'\right),
	\label{eq:tba_ac}
\end{align}
which holds in the fundamental strip. 
Note that a prime on the auxiliary parameters and the $\widetilde{\Yf}$-functions indicates the values of the parameters at the endpoint of the continuation, which will be related to the parameters at the starting point of the continuation later on.
Furthermore, we have introduced the objects $\partial_\theta\,\mathcal{S}^{a,a'}_{s,s'}(\theta):=-2\pi i\,\mathcal{K}^{a,a'}_{s,s'}(\theta)$ in eq.\ (\ref{eq:tba_ac}), which are called \textit{S-matrices} and which are specified in appendix \ref{app:smatrices}.
Which $\widetilde{\Yf}_{a,s}$-functions have crossing singularities depends on the paths of the auxiliary parameters during the analytic continuation and needs to be determined on a case-by-case basis.\par
Once we have established which solutions of eq.\ (\ref{eq:def_theta_tilde}) cross the integration contour, we can take the multi-Regge limit (\ref{eq:def_mrl}) at the endpoint of the continuation.
As in the Euclidean region, the contributions of the integrals are exponentially suppressed, and the equations for the $\widetilde{\Yf}'$-functions simplify,
\begin{equation}
	\log\widetilde{\Yf}'_{a,s}(\theta)\cong-|m_{a,s}|'\cosh\theta+C'_{a,s}+\sum\limits_{a',s'}\sum\limits_{i=1}^{n_{a',s'}} \mathrm{sign}(\mathrm{Im}\,\widetilde{\theta}_{a',s',i})\log \mathcal{S}^{a,a'}_{s,s'}(\theta-\tilde{\theta}_{a',s',i}+i\varphi_s'-i\varphi_{s'}').
	\label{eq:tba_ac_mrl}
\end{equation}
Similarly, the integrals in the $A_\mathrm{free}$-contribution to the remainder function at the endpoint of the continuation are negligible when going to the multi-Regge limit and we end up with a simple equation for $A'_\mathrm{free}$, which is determined by the configuration of crossing singularities and the TBA parameters,
\begin{align}
	A'_{\mathrm{free}} &= \sum\limits_{a,s}\frac{|m_{a,s}|'}{2\pi}\int\limits_{\mathbb{R}}\cosh\theta\log\left(1+\widetilde{\Yf}'_{a,s}(\theta)\right)+\sum\limits_{a,s}i|m_{a,s}|'\sum\limits_{i=1}^{n_{a,s}}\mathrm{sign}\left(\mathrm{Im}\,\widetilde{\theta}_{a,s,i}\right)\sinh\widetilde{\theta}_{a,s,i}\nonumber\\
	&\cong \sum\limits_{a,s}i|m_{a,s}|'\sum\limits_{i=1}^{n_{a,s}}\mathrm{sign}\left(\mathrm{Im}\,\widetilde{\theta}_{a,s,i}\right)\sinh\widetilde{\theta}_{a,s,i}.
	\label{eq:a_free_ac}
\end{align}
Importantly, the $A'_{\mathrm{free}}$-contribution no longer vanishes in the multi-Regge limit as in the Euclidean region, which ultimately gives rise to a non-trivial remainder function, as we will see later.\par
So far, the remainder function at the endpoint of the analytic continuation still explicitly depends on the auxiliary parameters at the endpoint of the continuation and the locations of the crossing singularities.
The latter can be determined analytically, by evaluating the corresponding $\widetilde{\Yf}_{a,s}$-functions at those points, which gives rise to the \textit{endpoint conditions} $-1=\widetilde{\Yf}_{a,s}(\tilde{\theta}_{a,s,i})$.
Assuming that the endpoints lie within the fundamental strip, the endpoint conditions read
\begin{equation}
	-1=\widetilde{\Yf}'_{a,s}(\tilde{\theta}_{a,s,i})=e^{-|m_{a,s}|'\cosh\tilde\theta_{a,s,i}+C'_{a,s}}\prod\limits_{a',s'}\prod\limits_{i=1}^{n_{a',s'}} \mathcal{S}^{a,a'}_{s,s'}\left(\theta-\tilde{\theta}_{a',s',i}+i\varphi'_s-i\varphi'_{s'}\right)^{\mathrm{sign}(\mathrm{Im}\,\widetilde{\theta}_{a',s',i})}.
	\label{eq:rev_endp_cond}
\end{equation}
These equations take the form of coordinate Bethe ansatz equations which can be solved for the endpoints $\widetilde{\theta}_{a,s,i}$.
This means that to every kinematic region of the multi-Regge limit we can associate a set of Bethe ansatz equations which characterize the pattern of crossing singularities of the corresponding analytic continuation.
Having fixed the endpoints $\tilde{\theta}_{a,s,i}$ of the crossing singularities, we still need to connect the TBA parameters $m_s'$, $\varphi'_s$ and $C_s'$ with the corresponding parameters at the starting point of the continuation.
Recall that as explained at the end of section \ref{sec:mrl_tba}, the relevant paths of analytic continuation for the cross ratios are such that at the endpoint 
\begin{equation}
	u_{1,s}' = u_{1,s},\quad u_{2/3,s}'=\pm u_{2/3,s}
	\label{eq:crs_signs}
\end{equation}
holds, with the choice of signs in eq.\ (\ref{eq:crs_signs}) depending on the kinematical region under consideration.
To determine the cross ratios $u'_{a,s}$ in the new kinematic region, we can use the relations between the cross ratios and the $\widetilde{\Yf}_{2,s}$-functions,
\begin{equation}
	u'_{1,s}=\frac{ {\Yf'}_{2,1}^{[2s+7]}}{1+{\Yf'}_{2,1}^{[2s+7]}},\quad u'_{2,s}=\frac{ {\Yf'}_{2,s}^{[s+4]}}{1+{\Yf'}_{2,s}^{[s+4]}},\quad u'_{3,s}=\frac{ {\Yf'}_{2,s}^{[s+6]}}{1+{\Yf'}_{2,s}^{[s+6]}},
	\label{eq:crs_ac}
\end{equation}
where ${\Yf'}^{[k]}_{a,s}:=\widetilde{\Yf}'_{a,s}(ik\frac{\pi}{4}-i\varphi_s)$ as before.
Using the relations (\ref{eq:crs_signs}) and (\ref{eq:crs_ac}) determines the auxiliary parameters $m'_s$ and $C'_s$ at the endpoint in terms of the parameters at the starting point of the continuation.
The value of these parameters, in general, differs from that at the starting point of the continuation.
In contrast, the parameters $\varphi'_s$ attain the same values after the analytic continuation as in (\ref{eq:mrl_aux}), i.e.\ $\varphi_s'=\varphi_s$, since these values are fixed by the condition that the ratio $\frac{u_{2,s}}{u_{3,s}}$ is finite in the multi-Regge limit (see \cite{Bartels:2012gq}), which holds at both the starting point and the endpoint of the continuation by eq.\ (\ref{eq:crs_signs}).\par
Then, finally, the remainder function in the Mandelstam region is given by
\begin{equation}
	R_n'=-\frac{\sqrt{\lambda}}{2\pi}\left(\Delta'(u_{a,s}')+A'_{\mathrm{per}}(m_s')+A'_{\mathrm{free}}(m_s', C_s')\right).
	\label{eq:def_rem_fct_endpoint}
\end{equation}
Recalling from section \ref{sec:mrl_tba} that the contributions of the remainder function cancel at the starting point of the analytic continuation, the remainder function can also be written in terms of the differences of the individual contributions,
\begin{equation}
	R_n'=-\frac{\sqrt{\lambda}}{2\pi}\Big(\left(\Delta'(u_{a,s}')-\Delta(u_{a,s})\right)+\left(A'_{\mathrm{per}}(m_s')-A_{\mathrm{per}}(m_s)\right)+\left(A'_{\mathrm{free}}(m_s', C_s')-A_{\mathrm{free}}(m_s, C_s)\right)\Big),
	\label{eq:def_rem_fct_endpoint_diff}
\end{equation}
which is simpler to evaluate in practice, as we will show when considering specific examples of the procedure described in this section for the nine-point amplitude in section \ref{sec:three_reggeon_bs}.\par
In this section, we have focused on the locations of the solutions of the equations $\widetilde{\Yf}_{a,s}(\theta)=-1$.
However, the integrands of the $\Yf$-system (\ref{eq:def_tba}) have additional singularities at the locations $\widetilde{\Yf}_{a,s}(\theta)=\infty$.
The discussion of this section holds for those locations, as well, the only difference being that all residues come with an additional minus sign (or, equivalently, that the S-matrix associated with such a crossing singularity is simply the inverse of a crossing singularity of the type $\widetilde{\Yf}_{a,s}=-1$ with the same indices $a$, $s$).
The locations of the solutions to $\widetilde{\Yf}_{a,s}(\theta)=\infty$  are not discussed in \cite{Bartels:2010ej, Bartels:2013dja, Bartels:2014ppa, Bartels:2014mka, Sprenger:2016jtx}, since such singularities were not observed to cross in the analytic continuations, which are performed numerically in those references.
As we will see in section \ref{sec:related_singularities}, these locations are closely tied to the locations of the solutions $\widetilde{\Yf}_{a,s}(\theta)=-1$ and play a central role in understanding the spectrum of possible crossing patterns.
\section{Singular points of the $\Yf$-system}
\label{sec:analysis_bae}
\subsection{Relations between singular points}
\label{sec:related_singularities}
After reviewing the general setup for the calculation of the remainder function in different kinematical regions of the multi-Regge limit, we now examine how the locations of the solutions $\widetilde{\Yf}_{a,s}(\theta)=-1$ and $\widetilde{\Yf}_{a,s}(\theta)=\infty$ in the complex $\theta$-plane are related.
The following observations were originally made in \cite{Dorey:1997rb} in the context of other TBAs and are adapted for the $\Yf$-system at hand in the following.
As explained in section \ref{sec:rev_ac}, the locations of these solutions can modify the $\Yf$-system equations during an analytic continuation to another kinematical region, which is a prerequisite of a non-trivial remainder function.
We begin our analysis in the Euclidean region of the multi-Regge limit (\ref{eq:mrl_aux}), i.e. before any analytic continuation of the $\Yf$-system is performed.
In the following, we will focus on singularities in the upper half-plane of the complex $\theta$-plane for simplicity -- all results hold symmetrically for singularities in the lower half-plane, however, a general discussion would unnecessarily complicate the formulas presented below.\par
\begin{figure}[thb]
	\centering
	\includegraphics{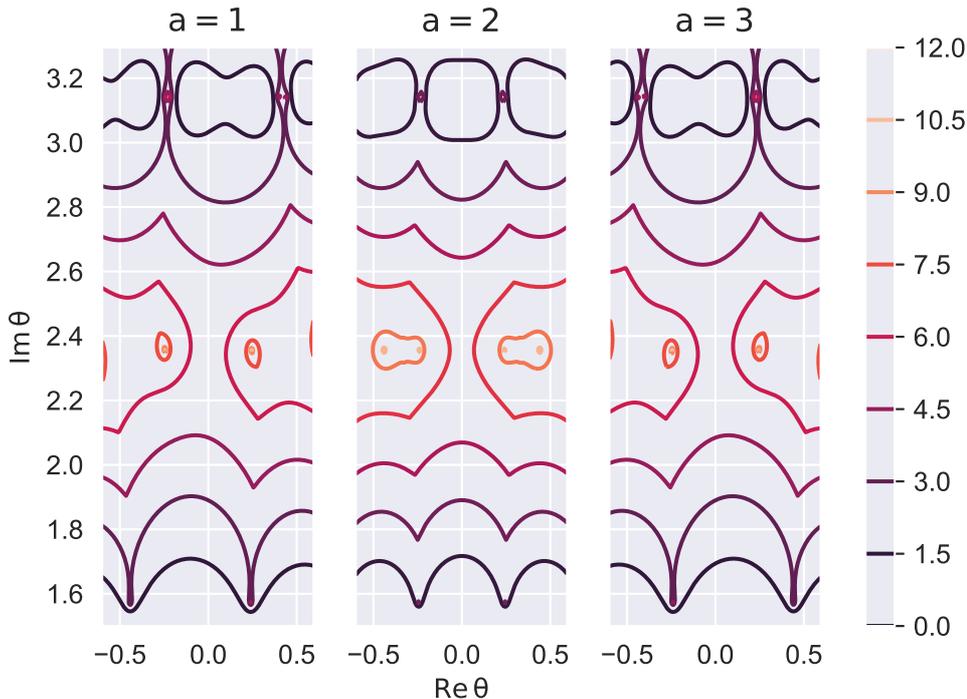}
	\caption{Contour plot of $|\log(1+\widetilde{\Yf}_{a,2}(\theta))|$ in the seven-point case for the three possible values of $a$. Points of dense contour lines correspond to singular points of the $\Yf$-system, with singular points in the vicinity of $\mathrm{Im}\,\theta=\frac{\pi}{2}$ and $\pi$ corresponding to singularities of the type $\widetilde{\Yf}_{a,2}=-1$ and singularities around $\mathrm{Im}\,\theta = 3\frac{\pi}{4} $ corresponding to singularities of the type $\widetilde{\Yf}_{a,2}=\infty$. Note that the locations satisfy the relations indicated in the middle column of table \ref{tab:rel_sing}.}
	\label{fig:contour_plot}
\end{figure}

As explained in section \ref{sec:review_tba}, when moving away from the real $\theta$-axis into the complex $\theta$-plane, the $\Yf$-system equations (\ref{eq:def_tba}) have to be modified due to singularities of the integration kernels, which cross the integration contour and whose residues have to be picked up.
Once the point in the $\theta$-plane on which we want to evaluate a given $\widetilde{\Yf}$-function is reached, we can drop the terms containing integrals, as those are negligible in the multi-Regge limit.
For example, assuming we move into the upper half-plane by increasing $\mathrm{Im}\,\theta$, the $\Yf$-system equations schematically read
\begin{equation}
	\widetilde{\Yf}_{a,s}(\theta)=e^{-\left|m_{a,s}\right|\cosh\theta+C_{a,s}}\cdot\prod\limits_{k=1}^n\left(1+\widetilde{\Yf}_{a_k,s_k}\left(\theta-n_k\cdot i\frac{\pi}{4}\right)\right)^{c_k},
	\label{eq:y_large_imag}
\end{equation}
where $n_k\in\mathbb{N}_0$, $c_k\in\left\{-1,1\right\}$.
In the following, we denote the exponential in eq.\ (\ref{eq:y_large_imag}) as the \textit{driving term}, and the remaining terms as \textit{residue contributions}.
The precise number of terms appearing on the right-hand side of eq.\ (\ref{eq:y_large_imag}) depends on $a$, $s$ and $\mathrm{Im}\,\theta$.
Since the parameters $\left|m_{a,s}\right|$ tend to infinity in the multi-Regge limit, the driving term of eq.\ (\ref{eq:y_large_imag}) will, depending on $\mathrm{Im}\,\theta$, either go to zero, go to infinity, or, at special points of $\theta$, be a pure phase.
The latter points are particularly interesting, as these are the points where solutions of $\widetilde{\Yf}_{a,s}(\theta)=-1$ may be found.
These solutions are found close to points at which $\mathrm{Re}\left(\cosh\theta\right)$ vanishes, which is the case when the imaginary part of $\theta$ is close to an odd multiple of $\frac{\pi}{2}$, see figure \ref{fig:contour_plot} for an example.\par
In contrast, the regions of the $\theta$-plane in which the driving term diverges in the multi-Regge limit are not necessarily the regions where solutions of $\widetilde{\Yf}_{a,s}(\theta)=\infty$ may be found, as we are interested in the points at which $\widetilde{\Yf}_{a,s}(\theta)=\infty$ holds already for large, but finite $\left|m_{a,s}\right|$.
To understand how such a singularity may arise, let us have a closer look at eq.\ (\ref{eq:y_large_imag}), which relates $\widetilde{\Yf}_{a,s}$-functions along the imaginary axis for the same value of $\mathrm{Re}\,\theta$.
As we have stated before, a singularity of the type $\widetilde{\Yf}_{a,s}=\infty$ cannot be generated by the driving term, since the parameters $|m_{a,s}|$ are assumed to be large but finite.
Hence, such a singularity can only arise through the residue contributions, namely when a $\widetilde{\Yf}_{a,s}$-function appearing in the denominator of eq.\ (\ref{eq:y_large_imag}) (i.e.\ a residue contribution with $c_k=-1$) equals $-1$.
This means that locations of the solutions to $\widetilde{\Yf}_{a,s}(\theta)=-1$ and $\widetilde{\Yf}_{a,s}(\theta)=\infty$ are closely related, and singularities of the form $\widetilde{\Yf}_{a,s}(\theta)=\infty$ can only appear if there is a singularity of the form $\widetilde{\Yf}_{a,s}(\theta)=-1$ at the same value of $\mathrm{Re}\,\theta$ at a smaller absolute value of $\mathrm{Im}\,\theta$, i.e. closer to the real axis.\par
\begin{figure}[h!]
	\centering
	\begin{tikzpicture}[node distance=1cm, auto]
		\tikzset{
		mynode/.style={rectangle, rounded corners, fill=blue!10, draw=blue!80, thick, text centered},
		myarrow/.style={draw=blue!80, thick, dashed, >=latex, shorten <=4pt, shorten >=4pt}	
		}
		
		\node[mynode] (upper row) {$\widetilde{\Yf}_{4-a_0,s_0}\left(\theta_0+i\frac{\pi}{2}\right)=-1$};
		\node[below=1.5cm of upper row] (dummy1) {};
		\node[mynode, left=of dummy1] (middle row left) {$\widetilde{\Yf}_{a_0-1, s_0}\left(\theta_0+i\frac{\pi}{4}\right)=\infty$};
		\node[mynode, right=of dummy1] (middle row right) {$\widetilde{\Yf}_{a_0+1, s_0}\left(\theta_0+i\frac{\pi}{4}\right)=\infty$};
		\node[mynode, below=1.5cm of dummy1] (lower row) {$\widetilde{\Yf}_{a_0,s_0}\left(\theta_0\right)=-1$};
		
		\draw[myarrow, <->, blue!80] (upper row.200)-- (middle row left.north);
		\draw[myarrow, <->, blue!80] (upper row.340) -- (middle row right.north);
		\draw[myarrow, <->, blue!80] (middle row left.south) -- (lower row.160);
		\draw[myarrow, <->, blue!80] (middle row right.south) -- (lower row.20);
	\end{tikzpicture}
\caption{Related singular points of the $\Yf$-system enforced by the recursion relations due to the occurrence of the singular point $\widetilde{\Yf}_{a_0,s_0}(\theta_0)=-1$.}\label{fig:diamond}
\end{figure}
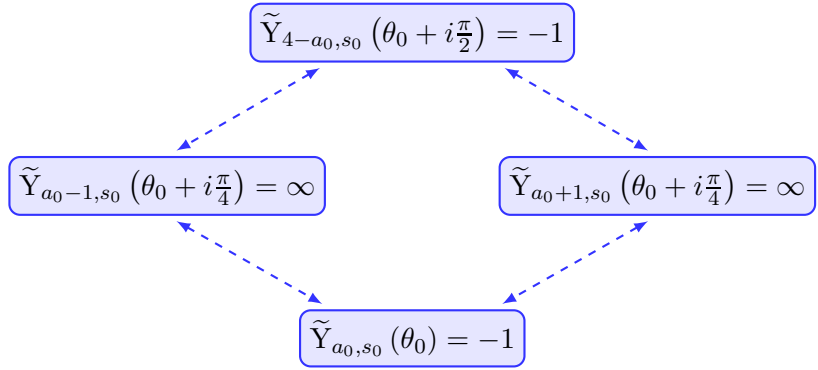

Indeed, assuming that $\widetilde{\Yf}_{a_0, s_0}(\theta_0)=-1$ with $\mathrm{Im}\,\theta_0>0$ holds, the recursion relations (\ref{eq:rec_rels_mrl}) immediately fix the value of several other $\widetilde{\Yf}_{a,s}$-functions, as summarized in table \ref{tab:rel_sing}.
In table \ref{tab:rel_sing}, we have only indicated the particular $\widetilde{\Yf}_{a,s}$-functions fixed by the occurrence of the singular point $\widetilde{\Yf}_{a_0,s_0}(\theta_0)=-1$, $\widetilde{\Yf}_{a,s}$-functions not specified in the table take some value not equal to $-1$, $0$ or $\infty$, which are not important in the following.
It should be noted that the pattern indicated in the table stops at $\theta_0+i\frac{\pi}{2}$, i.e. there are no singular points further away from the real axis induced by the singular point $\widetilde{\Yf}_{a_0, s_0}(\theta_0)=-1$.
\begin{table}
	\begin{center}
		{\renewcommand{\arraystretch}{1.5}
			\begin{tabular}[tbh]{|c||c|c|c|}
			\hline
			$\theta$ & $s=s_0-1$ & $s=s_0$ & $s=s_0+1$ \\
			\hline
			$\theta_0+i\frac{\pi}{2}$ & & $\widetilde{\Yf}_{4-a_0, s_0} = -1$ & $\widetilde{\Yf}_{a_0, s_0+1}=0$\\
			\hline
			$\theta_0+i\frac{\pi}{4}$ & & $\widetilde{\Yf}_{a_0\pm 1, s_0} = \infty$ & \\
			\hline
			$\theta_0$ & $\widetilde{\Yf}_{4-a_0, s_0-1}=0$ & $\widetilde{\Yf}_{a_0, s_0}=-1$ & \\
			\hline
		\end{tabular}}
	\end{center}
	\caption{Singular points fixed through the recursion relations (\ref{eq:rec_rels_mrl}) by the occurrence of a singular point $\widetilde{\Yf}_{a_0,s_0}(\theta_0)=-1$ with $\mathrm{Im}\,\theta_0>0$. Note that $\widetilde{\Yf}_{a,s}$-functions with $s\notin\{s_0-1, s_0, s_0+1\}$ are not affected by the singular point.}
	\label{tab:rel_sing}
\end{table}
As is visible in table \ref{tab:rel_sing}, the singularities for the index $s=s_0$ naturally form \textit{diamonds} as shown in figure \ref{fig:diamond}.
\par\noindent
Note that for the case $a_0=2$, the related singularities at $\theta_0+i\frac{\pi}{4}$ are indeed two singularities, namely $\widetilde{\Yf}_{1/3, s_0}(\theta_0+i\frac{\pi}{4})=\infty$.
For the cases $a_0=1/3$, only one related singularity, $\widetilde{\Yf}_{2, s_0}(\theta_0+i\frac{\pi}{4})=\infty$, needs to be taken into account due to the boundary conditions (\ref{eq:bdy_cond}).
\par
So far, our discussion was limited to the Euclidean region, i.e. before any analytic continuation to another kinematical region of the multi-Regge limit is carried out.
However, it is very important to note that the relations between the singular points of the $\Yf$-system directly follow from the recursion relations.
Since the recursion relations are preserved during any analytic continuation, so are the relations between the singular points.
This entails that, while the location of the solution $\widetilde{\Yf}_{a_0,s_0}(\theta_0)=-1$ may vary as we perform an analytic continuation of the parameters of the $\Yf$-system, the related singularities will move synchronously such that the relations indicated in table \ref{tab:rel_sing} hold throughout the continuation.
As a consequence, the possible modifications of the $\Yf$-system during an analytic continuation due to singular points of the $\widetilde{\Yf}_{a,s}$-functions crossing the integration contour are also constrained, as we will argue in the next section.
\subsection{Crossing diamonds}
\label{sec:crossing_diamonds}
In this section, we start analyzing what happens when one or several of the singular points of the diamond described in figure \ref{fig:diamond} cross the integration contour during any analytic continuation of the parameters of the TBA equations.
Note that, of course, not every analytic continuation will lead to singular points crossing the integration contour (and hence a non-trivial remainder function).
Rather, we are interested in the question which patterns of crossing singularities are possible \textbf{if} crossings occur in a given analytic continuation.
Furthermore, we focus on crossing singularities belonging to a single diamond for simplicity. 
Of course, in a given analytic continuation, an arbitrary number of diamonds can have crossing singularities.
However, in this section we are only interested in which S-matrices we need to take into account at the endpoint of any analytic continuation and from that perspective the different diamonds are independent from each other and can be analyzed separately.
Lastly, as in section \ref{sec:related_singularities}, we describe all formulas for singular points in the positive half-plane of the complex $\theta$-plane crossing into the negative half-plane during an analytic continuation.
The final results of this section, however, hold for singularities crossing in either direction.
\subsubsection{Case 1: $-\frac{\pi}{4}\leq\mathrm{Im}\,\theta_0'\leq 0$}
\begin{figure}[t]
	\centering
	\includegraphics[scale=.7]{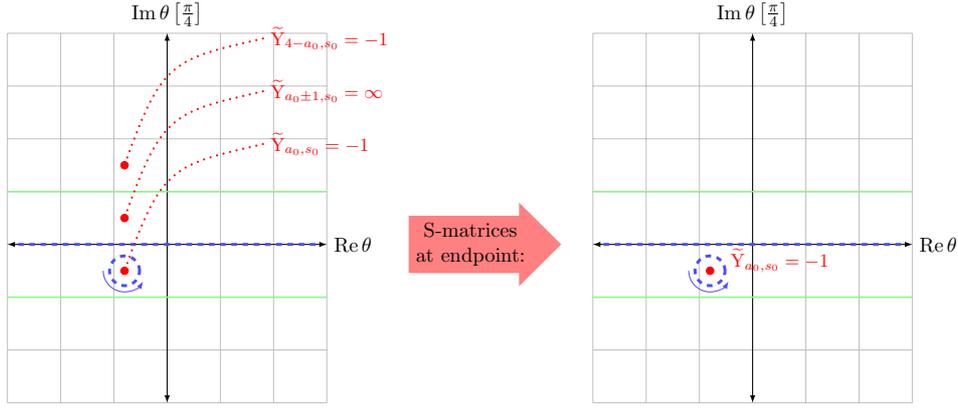}
	\caption{Example of an analytic continuation in which only the first singular point of the diamond shown in figure \ref{fig:diamond} crosses the integration contour, while the other singular points remain above the integration contour indicated by the blue dashed line. The blue arrow indicates the orientation that the residue contribution of the crossing singularity is picked up with. The green solid line indicates the region $-\frac{\pi}{4}\leq\mathrm{Im}\,\theta\leq\frac{\pi}{4}$. Note that the locations of the singularities $\widetilde{\Yf}'_{a_0\pm 1, s_0}=\infty$ are within this region, but have not crossed the integration contour by assumption. At the endpoint, one S-matrix corresponding to the crossing singularity has to be taken into account.}
	\label{fig:crossing_case1}
\end{figure}

Due to the ordering of the singularities shown in figure \ref{fig:diamond}, the first singular point to cross the integration contour necessarily is the location of the solution to $\widetilde{\Yf}'_{a_0,s_0}(\theta_0')=-1$, where the prime indicates that the location of this singularity has moved from its original point $\theta_0$ during the continuation.
As long as $-\frac{\pi}{4}\leq\mathrm{Im}\,\theta_0'<0$ holds throughout the continuation, all related singularities shown in figure \ref{fig:diamond} do not cross the integration contour.
Hence, at the endpoint of the continuation we would have to take into account one crossing singularity of the type $\widetilde{\Yf}'_{a_0, s_0}(\theta_0')=-1$, with $\mathrm{Im}\,\theta_0'>-\frac{\pi}{4}$, as depicted in figure \ref{fig:crossing_case1}.
This case was already covered in section \ref{sec:rev_ac}.
\subsubsection{Case 2: $-\frac{\pi}{2}\leq\mathrm{Im}\,\theta_0'<-\frac{\pi}{4}$}
\begin{figure}[t]
	\centering
	\includegraphics[scale=.7]{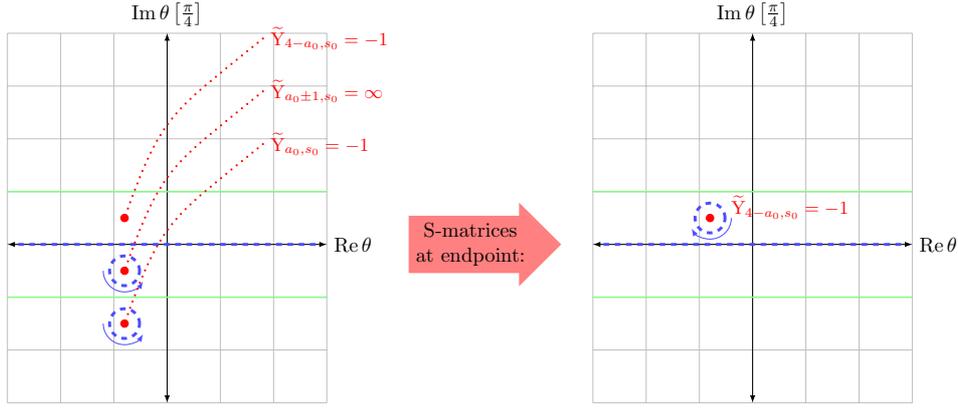}
	\caption{Example of an analytic continuation in which the two lowest layers of the diamond shown in figure \ref{fig:diamond} cross the integration contour, while the uppermost singular point of the diamond remains above the integration contour indicated by the dashed blue line. The blue arrow indicates the orientation that the residue contributions of the crossing singularities are picked up with. The green solid line indicates the region $-\frac{\pi}{4}\leq\mathrm{Im}\,\theta\leq\frac{\pi}{4}$. Note that the locations of the singularity $\widetilde{\Yf}'_{4-a_0, s_0}=-1$ are within this region, but have not crossed the integration contour by assumption. At the endpoint, despite several crossing singularities, only one S-matrix contribution corresponding to an effective singularity of the type $\widetilde{\Yf}'_{4-a_0,s_0}=-1$ has to be taken into account.}
	\label{fig:crossing_case2}
\end{figure}

Let us now assume that $-\frac{\pi}{2}\leq\mathrm{Im}\,\theta_0'<-\frac{\pi}{4}$.
In this case, the two related singularities $\widetilde{\Yf}'_{a_0\pm 1, s_0}\left(\theta_0'+i\frac{\pi}{4}\right)=\infty$ have also crossed the integration contour, and the corresponding S-matrices have to be taken into account as described at the end of section \ref{sec:rev_ac}.
Interestingly, using the relations
\begin{align}
	&S_1\left(x+i\frac{\pi}{4}\right)S_1\left(x-i\frac{\pi}{4}\right)=-S_2(x),\nonumber\\
	&S_2\left(x+i\frac{\pi}{4}\right)S_2\left(x-i\frac{\pi}{4}\right)=-S_1(x)^2,\nonumber\\
	&S_3\left(x+i\frac{\pi}{4}\right)=-S_3\left(x-i\frac{\pi}{4}\right)\label{eq:rel_smat}
\end{align}
between the basic S-matrices (see section \ref{app:smatrices}), the S-matrices of the crossed singularities can be combined into a single S-matrix,
\begin{align}
	&\frac{\mathcal{S}_{s,s_0}^{a,a_0}(\theta-\theta_0'+i\varphi_s-i\varphi_{s_0})}{\mathcal{S}_{s,s_0}^{a,a_0+1}(\theta-(\theta_0'+i\frac{\pi}{4})+i\varphi_s-i\varphi_{s_0})\,\mathcal{S}_{s,s_0}^{a,a_0-1}(\theta-(\theta_0'+i\frac{\pi}{4})+i\varphi_s-i\varphi_{s_0})}\nonumber\\
	&\quad\quad\quad=-\mathcal{S}_{s,s_0}^{a,4-a_0}\left(\theta-\left(\theta_0'+i\frac{\pi}{2}\right)+i\varphi_s-i\varphi_{s_0}\right),
	\label{eq:effective_smatrix}
\end{align}
so that the crossing pattern formally looks like the crossing of a singularity of the type $\widetilde{\Yf}'_{a_0, s_0}(\theta_0'+ i \frac{\pi}{2})=-1$, which has crossed the integration contour from the negative half-plane into the positive half-plane and is located at $\theta_0'+i\frac{\pi}{2}$.\footnote{We comment on the appearance of the additional minus sign and its implications for the remainder function in appendix \ref{sec:app_bae}.}
This is illustrated in figure \ref{fig:crossing_case2}.
Note that under the assumption $-\frac{\pi}{2}<\mathrm{Im}\,\theta_0'<-\frac{\pi}{4}$, the imaginary part of this \textit{effective singularity} is smaller than $\frac{\pi}{4}$.
Thus, as in the first case, at the endpoint of the continuation we have to take into account a single crossing singularity of the type $\widetilde{\Yf}'_{4-a_0, s_0}(\theta_0'')=-1$, with $0<\mathrm{Im}\,\theta_0''<\frac{\pi}{4}$.
\par\noindent
Recall from section \ref{sec:rev_ac} that crossing singularities modify both the $\Yf$-system and the $A_{\mathrm{free}}$-contribution to the remainder function (see eq.\ (\ref{eq:a_free_ac})).
Due to the relation
\begin{align}
	A'_{\mathrm{free}}&=\dots\,-i|m_{a_0,s_0}|'\sinh\theta_0'+i|m_{a_0+1,s_0}|'\sinh\left(\theta_0'+i\frac{\pi}{4}\right)+i|m_{a_0-1,s_0}|'\sinh\left(\theta_0'+i\frac{\pi}{4}\right)\nonumber\\
	&=\dots\,-i|m_{4-a_0,s_0}|'\sinh\left(\theta_0'+i\frac{\pi}{2}\right),
	\label{eq:effective_a_free}
\end{align}
where the dots indicate the integral contributions which are irrelevant at the endpoint of the continuation, the same effective crossing singularity obtained for the $\Yf$-system equations also describes the $A_{\mathrm{free}}$-contribution of the crossing singularities to the remainder function.\footnote{One subtlety to note here is that this relation is satisfied differently for the possible values of $a_0$: Recall that $|m_{2,s_0}|'=\sqrt{2}|m_{1/3,s_0}|'$ (see eq.\ (\ref{eq:def_params_tba})). However, for the case $a_0=2$ two singularities, namely $\widetilde{\Yf}'_{1/3, s_0}(\theta_0'+i\frac{\pi}{4})=\infty$ contribute in eq.\ (\ref{eq:effective_a_free}), which compensate for that difference in the definition of the parameter $|m_{a,s}|$. For the cases $a_0=1/3$, only one additional singularity, namely $\widetilde{\Yf}'_{2,s_0}(\theta_0'+i\frac{\pi}{4})=\infty$ contributes due to the boundary conditions (\ref{eq:bdy_cond}).}
\subsubsection{Case 3: $\mathrm{Im}\,\theta_0'<-\frac{\pi}{2}$}
\begin{figure}[t]
	\centering
	\includegraphics[scale=.7]{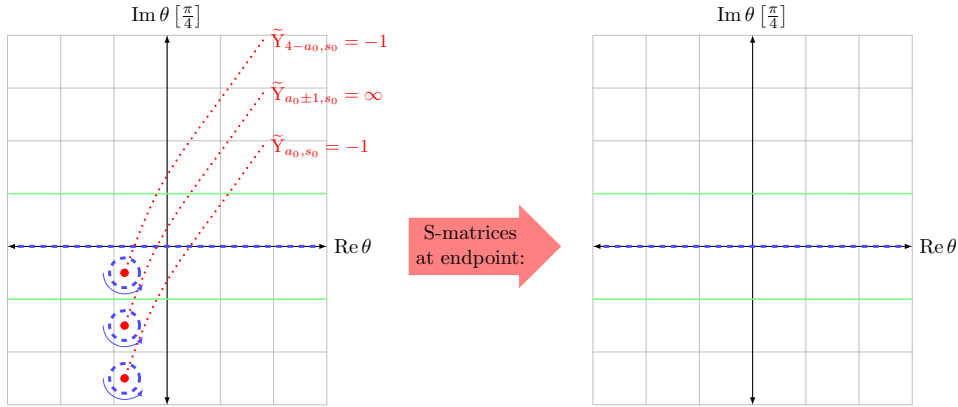}
	\caption{Example of an analytic continuation in which all singularities of the diamond shown in figure \ref{fig:diamond} cross the integration contour indicated by the dashed blue line. The blue arrow indicates the orientation that the residue contributions of the crossing singularities are picked up with. The green solid line indicates the region $-\frac{\pi}{4}\leq\mathrm{Im}\,\theta\leq\frac{\pi}{4}$. At the endpoint, the S-matrix contributions of the crossing singularities cancel and no S-matrices have to be taken into account.}
	\label{fig:crossing_case3}
\end{figure}

Lastly, let us assume that $\mathrm{Im}\,\theta_0' <-\frac{\pi}{2}$.
In this case, all related singularities of figure \ref{fig:diamond} have also crossed the integration contour and remarkably the corresponding S-matrices cancel,
\begin{equation}
	\frac{\mathcal{S}_{s,s_0}^{a,a_0}(\theta-\theta_0'+i\varphi_s-i\varphi_{s_0})\,\mathcal{S}_{s,s_0}^{a,4-a_0}(\theta-(\theta_0'+i\frac{\pi}{2})+i\varphi_s-i\varphi_{s_0})}{\mathcal{S}_{s,s_0}^{a,a_0+1}(\theta-(\theta_0'+i\frac{\pi}{4})+i\varphi_s-i\varphi_{s_0})\,\mathcal{S}_{s,s_0}^{a,a_0-1}(\theta-(\theta_0'+i\frac{\pi}{4})+i\varphi_s-i\varphi_{s_0})}=1.
	\label{eq:effective_full_crossing}
\end{equation}
This means that, even though singularities of the $\Yf$-system have crossed the integration contour, the corresponding contributions to the $\Yf$-system (and similarly to $A_{\mathrm{free}}$) vanish and can therefore be neglected in the calculation of the remainder function.
This is depicted in figure \ref{fig:crossing_case3}.
Summarizing our discussion so far:
\begin{center}
\begin{tikzpicture}
	\node[draw=blue!80, fill=blue!10, very thick, rectangle, rounded corners, inner sep=10pt, inner ysep=10pt] (resbox){%
		\begin{minipage}{.95\textwidth}
			We have shown that the locations of the singularities of the $\Yf$-system (\ref{eq:def_tba}) of the type $\widetilde{\Yf}_{a,s}=-1$ and of the type $\widetilde{\Yf}_{a,s}=\infty$ are connected through the recursion relations (\ref{eq:rec_rels_mrl}) as summarized in figure \ref{fig:diamond} and move synchronously when the $\Yf$-system is analytically continued into another kinematical region. The relations between the singularities of the $\Yf$-system imply that the most general pattern of crossing singularities consists of an arbitrary number of crossing singularities of the type $\widetilde{\Yf}_{a,s}=-1$, whose location at the endpoint of the continuation lies in the region 
		\begin{equation}
			-\frac{\pi}{4}\leq\mathrm{Im}\,\theta_0'\leq\frac{\pi}{4}.
		\nonumber
		\end{equation}
		\end{minipage}
	};
\end{tikzpicture}
\end{center}
\par\noindent
The latter aspect is important because in this region, we can solve the corresponding Bethe ansatz equations, as explained in the next section. 
\subsection{Solving the Bethe ansatz}
\label{sec:constraints}
In the previous section, we have argued that the most general pattern of crossing singularities in any analytic continuation of the $\Yf$-system from the multi-Regge limit in the Euclidean region to another kinematical region consists of an arbitrary number of crossing singularities of the type $\widetilde{\Yf}'_{a, s}(\tilde{\theta}_{a,s,i})=-1$, where $i=1,\dots,n_{a,s}$ parametrizes the crossing singularities for given indices $a$, $s$, whose endpoints lie in the strip $-\frac{\pi}{4}\leq \mathrm{Im}\,\tilde{\theta}_{a,s,i}\leq \frac{\pi}{4}$.
In the following, we will also distinguish between the number $n^{(+)}_{a,s}$ of crossing singularities with endpoints in the positive half-plane and the number $n^{(-)}_{a,s}$ of crossing singularities with endpoints in the negative half-plane for given values of the indices $a$ and $s$ (these numbers are related to the parametrization used so far by $n_{a,s}=n^{(+)}_{a,s}+n^{(-)}_{a,s}$).
As explained in section \ref{sec:rev_ac}, for each configuration of crossing singularities, there is an associated set of Bethe ansatz equations, which determine the endpoints of the crossing singularities exactly (see eq.\ \eqref{eq:rev_endp_cond}).
While the results of the previous section still allow a considerable number of possible Bethe ansatz equations, it turns out that there is a single physically-relevant solution in the multi-Regge limit.
The general idea in finding this solution is to study the endpoint conditions (\ref{eq:rev_endp_cond}) for the most general Bethe ansatz.
This by itself does not uniquely fix the endpoints of the crossing singularities.
However, the multi-Regge limit imposes further constraints on the behavior of the cross ratios at the endpoint of the continuation, which are also affected by the crossing singularities.
This additional input fixes the endpoints of all crossing singularities.
Since the derivation of this result is rather technical, we only state the main result here and provide the details on the derivation in appendix \ref{sec:app_bae}:
\begin{center}
\begin{tikzpicture}
	\node[draw=blue!80, fill=blue!10, very thick, rectangle, rounded corners, inner sep=10pt, inner ysep=10pt] (resbox){%
		\begin{minipage}{.95\textwidth}
			The most general configuration of crossing singularities allowed in the multi-Regge limit consists of a number $n_{1/3,s}^{(+)}$ crossing singularities of the form $\widetilde{\Yf}'_{1/3,s}(\tilde{\theta}^{(+)}_{1/3,s,i})=-1$ with endpoints in the positive half-plane and $n_{1/3, s}^{(-)}$ crossing singularities of the form $\widetilde{\Yf}'_{1/3,s}(\tilde{\theta}^{(-)}_{1/3, s, i})=-1$ with endpoints in the negative half-plane.
			In particular, no crossing singularities of the type $\widetilde{\Yf}'_{2,s}(\tilde{\theta}^{(+/-)}_{2,s,i})=-1$ are allowed.
			Furthermore, the endpoints of the crossing singularities are uniquely fixed to
			\begin{equation}
				\tilde{\theta}^{(+)}_{a,s,i}=i\frac{\pi}{4},\quad \tilde{\theta}^{(-)}_{a,s,i}=-i\frac{\pi}{4}.\nonumber
			\end{equation}
		\end{minipage}
	};
\end{tikzpicture}
\end{center}
\par\noindent
This simple solution of the Bethe ansatz equations allows us to extract the possible BFKL eigenvalues for a general $n$-point amplitude in section \ref{sec:rem_fct}.
\section{Remainder functions from the Bethe ansatz and BFKL eigenvalues}
\label{sec:rem_fct}
Having found the possible patterns of crossing singularities in section \ref{sec:constraints}, we proceed by calculating the remainder function for these configurations.
Before we delve into the calculation, however, let us briefly discuss which parts of the remainder function can actually be determined based on the solution of the Bethe ansatz alone.
Recall that the contribution $\Delta$ to the remainder function (see eq.\ (\ref{eq:def_rem_fct})) is a function of the cross ratios $u_{a,s}$ of transcendentality two.
During an analytic continuation of the cross ratios, this term may pick up discontinuities $\sim 2\pi i \log u_{a,s}$, which result in phases appearing in the remainder function at the endpoint of the continuation (see \cite{Bartels:2010ej, Bartels:2013dja, Bartels:2014mka} for explicit examples in the six- and seven-point case).
These phases, however, can only be calculated by explicitly specifying the path of continuation.
Since we are only looking at the endpoint of any analytic continuation, we cannot fully determine the phase of the remainder function using this approach.
The same applies to constants, which, for example, may arise from double discontinuities of terms in the $\Delta$-contribution to the remainder function.
Hence, in the following, we also do not calculate phases appearing in the other components of the remainder function, but mention where such terms are neglected.
This is not a limitation of our results, since the BFKL eigenvalues can still be fully calculated as described below.\par
Based on the results of the previous section, we start with the most general crossing pattern, i.e.\ an arbitrary number $n_{1/3, s}^{(+/-)}$ of crossing singularities of the type $\widetilde{\Yf}_{1/3,s}=-1$ with endpoints $\tilde{\theta}_{1/3, s, i}^{(+/-)}=\pm i \frac{\pi}{4}$.
Note that the remainder function treats crossing singularities of the two cases $a=1/3$ identically,\footnote{This is not true for the $\Yf$-system in general, where the terms $\gamma_s$ explicitly break this symmetry (see eq.\ (\ref{eq:ysys_app})). However, the remainder function is determined based on the cross ratios, i.e.\ the functions $\widetilde{\Yf}_{2,s}$, and the contribution $A_\mathrm{free}$, which are symmetric in $\widetilde{\Yf}_{1,s}\leftrightarrow\widetilde{\Yf}_{3,s}$.} so that the most general crossing pattern is fully specified by two values for each value of the index $s$.\par
We begin by calculating the parameters $\varepsilon_s'$ and $w_s'$ introduced in eq.\ (\ref{eq:def_mrl_param}) at the endpoint of the continuation,
\begin{align}
	\log\varepsilon_s&=\frac{1}{2}\log\left(u_{2,n-4-s}\,u_{3,n-4-s}\right)=\frac{1}{2}\log\left(u'_{2,n-4-s}\,u'_{3,n-4-s}\right)\cong\frac{1}{2}\log\left(\widetilde{\Yf}_{2,s}\left(i\frac{\pi}{4}\right)\widetilde{\Yf}_{2,s}\left(-i\frac{\pi}{4}\right)\right)\nonumber\\
	&=\log\varepsilon'+\frac{1}{4}\left(n_{1/3,s}^{(+)}+n_{1/3,s}^{(-)}\right)\log\frac{S_2\left(i\frac{\pi}{2}\right)}{S_2\left(-i\frac{\pi}{2}\right)}-\frac{1}{2}n_{1/3,s-1}^{(+)}\log\frac{S_1\left(i\frac{\pi}{4}\right)}{S_1\left(-3i\frac{\pi}{4}\right)}\nonumber\\
	&\quad\quad-\frac{1}{2}n_{1/3, s+1}^{(-)}\log\frac{S_1\left(3i\frac{\pi}{4}\right)}{S_1\left(-i\frac{\pi}{4}\right)},\label{eq:rel_eps_aper_smat}
\end{align}
which follows immediately from eq.\ (\ref{eq:general_y2_crs}) and the most general allowed crossing pattern (see section \ref{sec:constraints}).
Evaluating the S-matrices leads to the following relations between the parameters $\varepsilon_s$ and $w_s$ at the starting point and the endpoint of the analytic continuation
\begin{align}
	\log\varepsilon'_s=\log\varepsilon_s+\log(1+\sqrt{2})\left(n_{1/3, s+1}^{(-)}+n_{1/3, s-1}^{(+)}-n_{1/3,s}^{(+)}-n_{1/3, s}^{(-)}\right)+\dots,\nonumber\\
	\log w'_s=\log w_s+\log(1+\sqrt{2})\left(n_{1/3, s+1}^{(+)}-n_{1/3, s-1}^{(-)}+n_{1/3,s}^{(-)}-n_{1/3, s}^{(+)}\right)+\dots,
	\label{eq:rel_aper}
\end{align}
where the relation for the parameter $w_s$ was derived analogously to eq.\ (\ref{eq:rel_eps_aper_smat}).
The dots in eq.\ (\ref{eq:rel_aper}) indicate phases, which we cannot fix from the Bethe ansatz alone, as mentioned above.\par
Using these relations, we can evaluate the different contributions to the remainder function.
$A_{\mathrm{free}}'$ naturally splits into a sum over the crossing singularities for the different values of the index $s$ (see eq.\ (\ref{eq:a_free_ac})),
\begin{align}
	A_{\mathrm{free}}'-A_{\mathrm{free}}&=\sum\limits_{a,s} i |m_{a,s}|'\sum\limits_{i=1}^{n_{a,s,i}}\mathrm{sign}\left(\mathrm{Im}\,\tilde{\theta}_{a,s,i}\right)\sinh\tilde{\theta}_{a,s,i}=\sum\limits_s -\frac{1}{\sqrt{2}}|m_s|'\left(n_{1/3,s}^{(+)}+n_{1,3,s}^{(-)}\right)\nonumber\\
	&=\sum\limits_s-\frac{1}{\sqrt{2}}\left(n_{1/3,s}^{(+)}+n_{1,3,s}^{(-)}\right)\sqrt{\log^2\varepsilon'_s+\log^2 w'_s}\cong\sum\limits_s\frac{1}{\sqrt{2}}\left(n_{1/3,s}^{(+)}+n_{1,3,s}^{(-)}\right)\log\varepsilon'_s\nonumber\\
	&=\sum\limits_s\frac{1}{\sqrt{2}}\left(n_{1/3,s}^{(+)}+n_{1,3,s}^{(-)}\right)\log\varepsilon_s+\dots,
	\label{eq:afree_endp_rem}
\end{align}
where we have used the endpoints of the crossing singularities in the second step and used the relation $\sqrt{\log^2\varepsilon_s'+\log^2 w'_s}\cong -\log\varepsilon_s'$, which holds in the multi-Regge limit since $\varepsilon_s'$ becomes small, while $w'_s$ remains finite.
The dots in the last step of eq.\ (\ref{eq:afree_endp_rem}) indicate constants, which arise due to the difference between $\varepsilon$ and $\varepsilon'$ (see eq.\ (\ref{eq:rel_aper})), but which we ignore as explained above.\par
The contribution $A_{\mathrm{per}}$ is more involved.
In particular, it is not obvious that this contribution can be written as a sum over the index $s$.
However, in appendix \ref{app:aper}, we show that this is indeed the case and that 
\begin{align}
	A'_{\mathrm{per}}&(\varepsilon'_s, w'_s)-A_{\mathrm{per}}(\varepsilon_s, w_s)=\nonumber\\
	&\quad\frac{1}{2}\log(1+\sqrt{2})\sum\limits_s\left(-\log\varepsilon_s (n_{1/3, s}^{(+)}+n_{1/3,s}^{(-)})+\log w_s(n_{1/3,s}^{(-)}-n_{1/3,s}^{(+)})\right)+\dots
	\label{eq:delta_aper_mt}
\end{align}
holds, where the dots indicate phases, which we ignore.
Lastly, as explained above, the contribution $\Delta'-\Delta$ to the remainder function consists of phases and constants only, and will therefore be ignored in the following, as well.
Combining the three contributions, we find that the remainder function has a very simple form and reads
\begin{align}
	\left.R_n\right|_{\mathrm{MRL}}=\frac{\sqrt{\lambda}}{2\pi} & \left[ \left(\frac{1}{2}\log(1+\sqrt{2})-\frac{1}{\sqrt{2}}\right)\sum\limits_s(n_{1/3,s}^{(+)}+n_{1/3,s}^{(-)})\log\varepsilon_s\right. \nonumber\\
	& \quad\quad \left.  +\frac{1}{2}\log(1+\sqrt{2})\sum\limits_s(n_{1/3,s}^{(+)}-n_{1/3,s}^{(-)})\log w_s \right].
	\label{eq:rem_fct_endp}
\end{align}
Focusing on the terms $\sim\log\varepsilon_s$, we can spell out this result in terms of Mandelstam variables using the kinematical identities 
\begin{equation}
	\varepsilon_{n-4-s} = u_{2,s}\,u_{3,s} = (1-u_{1,s})\,\tilde{u}_{2,s}\,\tilde{u}_{3,s}
	\label{eq:eps_rem_fct}
\end{equation}
and $1-u_{1,s}\sim s_{s+1}^{-1}$ (see e.g.\ \cite{Bartels:2012gq}), which shows that the amplitude (\ref{eq:def_amplitude}), being proportional to the exponential of the remainder function, has a power-law like Regge behavior. 
This justifies our interpretation of the coefficients of $\log\varepsilon_s$ as the BFKL eigenvalue in this particular channel.
Note that the case $n_{1/3,s}^{(+)}=n_{1/3,s}^{(-)}=1$ for a given channel results in the two-Reggeon BFKL eigenvalue $e_2=\log(1+\sqrt{2})-\sqrt{2}$ and is the crossing pattern observed in the six-point remainder function for the Mandelstam region $(--)$ \cite{Bartels:2010ej, Bartels:2013dja}.\par
At this point, we have used all information of the solution of the most general crossing pattern.
However, we can constrain the remainder function even further by looking at target-projectile symmetry.
This symmetry exchanges the two incoming particles and acts as
\begin{equation}
	u_{1,s}\leftrightarrow u_{1,n-4-s},\quad u_{2,s}\leftrightarrow u_{3,n-4-s}
	\label{eq:tp_sym_crs}
\end{equation}
on the cross ratios (see, for example, \cite{Bartels:2014mka}), which implies that it acts as
\begin{equation}
	\varepsilon_s \leftrightarrow \varepsilon_{n-4-s},\quad w_s\leftrightarrow \frac{1}{w_{n-4-s}}
	\label{eq:tp_symmetry}
\end{equation}
on our kinematic parameters.
Importantly, target-projectile symmetry relates the remainder function in the regions $(s_1\, s_2\, \cdots\, s_n) \leftrightarrow (s_n\, s_{n-1}\, \cdots\, s_1)$, where $s_i\in\{+, -\}$, such that the structure of the remainder function is the same, but it is described in terms of different kinematic variables in the different regions,\footnote{As the simplest example for this effect of target-projectile symmetry, in the seven-point case it holds that $R_{7, (--+)}(u_{1,1}, u_{2,1}, u_{3,1})=R_{7, (+--)}(u_{1,2}, u_{2,2}, u_{3,2})$.} unless the region is target-projectile symmetric in which case the remainder function must be invariant under the symmetry.\par
As a first application of target-projectile symmetry, let us consider the regions in which Regge cuts are expected to contribute to the remainder function for the first time.
Regge theory predicts that when increasing the number $n$ of gluons under consideration, the first time a new Reggeon bound state contributes to the remainder function, it does so in a kinematic region which is target-projectile symmetric \cite{Lipatov:2009nt}.\footnote{For example, the three-Reggeon bound state is expected to contribute to the remainder function in the $(-++-)$-region of the eight-point amplitude \cite{Lipatov:2009nt, Bartels:2020twc} and the four-Reggeon bound state is expected to appear in the $(-+--+-)$-region of the ten-point amplitude \cite{Lipatov:2009nt}, both of which are target-projectile symmetric. Also note that in both cases, the central channel in which the new bound state appears is mapped to itself under target-projectile symmetry.}
Therefore, as long as we are only interested in the spectrum of possible BFKL eigenvalues, we can limit ourselves to kinematic regions which are invariant under target-projectile symmetry.
To be more specific, we can always choose $n=4k+2$ ($k\in\mathbb{N}$).
In this case, Regge theory predicts that there are Mandelstam regions in which the $m$-Reggeon bound state ($m=2,\dots, 2k$) propagates in the central channel with index $s=2k-1$ and which are symmetric under target-projectile symmetry.
This central channel is mapped to itself under target-projectile symmetry (\ref{eq:tp_symmetry}).
Therefore, target-projectile symmetry requires $n_{1/3,2k-1}^{(+)}=n_{1/3,2k-1}^{(-)}$ to hold (cf.\ eq.\ (\ref{eq:eps_rem_fct})).
Using this constraint, we see that all possible BFKL eigenvalues are multiples of the six-point BFKL eigenvalue $e_2$,
\begin{equation}
	\left(\frac{1}{2}\log(1+\sqrt{2})-\frac{1}{\sqrt{2}}\right)(n_{1/3,2k-1}^{(+)}+n_{1/3,2k-1}^{(-)})=\left(\log(1+\sqrt{2})-\sqrt{2}\right)\cdot n_{1/3,2k-1}^{(+)},
	\label{eq:spectrum_bfkl}
\end{equation}
which is the main result of this paper.
Note that specifying the number of gluons to $n=4k+2$ does not limit this result -- if the remainder function is consistent with Regge theory for all $n$, then the BFKL eigenvalue of the $m$-Reggeon bound state must be the same, independent of the number of external gluons.
We only use the specific case $n=4k+2$, since it allows us to leverage target-projectile symmetry.\footnote{Another reason to choose this case is the subtlety of the additional contribution $A_{\mathrm{extra}}$ to the remainder function for the case $n=4k$ mentioned in section \ref{sec:review_tba}. Note, however, that our analysis shows that, if the remainder function is compatible with Regge theory for all $n$, this additional piece cannot lead to different BFKL eigenvalues.}\par
At this point, we have determined all possible BFKL eigenvalues.
However, we cannot conclude yet that the BFKL eigenvalue of a $m$-Reggeon bound state always shows up with the same pattern of crossing singularities, since eq.\ (\ref{eq:rem_fct_endp}) allows several combinations of crossing singularities resulting in the same BFKL eigenvalue.
To do that, we need to use target-projectile symmetry again.
Focusing on a particular channel $s_0$, the contribution of this channel to the remainder function would be mapped to the channel $n-4-s_0$ under target-projectile symmetry,
\begin{align}
	& (n_{1/3, s_0}^{(+)}+n_{1/3, s_0}^{(-)})\log\varepsilon_{s_0} + (n_{1/3, s_0}^{(+)}-n_{1/3, s_0}^{(-)})\log w_{s_0}\nonumber\\
	& \quad\quad\longleftrightarrow\quad (n_{1/3, s_0}^{(+)}+n_{1/3, s_0}^{(-)})\log\varepsilon_{n-4-s_0} - (n_{1/3, s_0}^{(+)}-n_{1/3, s_0}^{(-)})\log w_{n-4-s_0},
	\label{eq:ex_tp_sym}
\end{align}
where we have dropped the overall factors appearing in eq.\ (\ref{eq:rem_fct_endp}).
As mentioned above, applying target-projectile symmetry may change the kinematic variables in which a remainder function is expressed, but it cannot change the kinematic dependence itself.
Accordingly, the prefactor of $\log w_{n-4-s_0}$ in eq.\ (\ref{eq:ex_tp_sym}) must agree with that of $\log w_{s_0}$ before applying target-projectile symmetry.
Looking at eq.\ (\ref{eq:ex_tp_sym}), we see that the kinematic dependence is unchanged only if $n_{1/3, s_0}^{(+)}=n_{1/3, s_0}^{(-)}$, since otherwise the term $\sim\log w_{s_0}$ spoils this property.
This constrains the crossing patterns which are consistent with the multi-Regge limit further and we can conclude that the BFKL eigenvalue $k\cdot e_2$ is uniquely described by the crossing pattern $n_{1/3, s}^{(+)}=n_{1/3, s}^{(-)}=k$.
Let us summarize the results of this section, before discussing a specific example in the following section:
\begin{center}
\begin{tikzpicture}
	\node[draw=blue!80, fill=blue!10, very thick, rectangle, rounded corners, inner sep=10pt, inner ysep=10pt] (resbox){%
		\begin{minipage}{.95\textwidth}
			The remainder function of the most general pattern of crossing singularities allowed by the multi-Regge limit as derived in section \ref{sec:constraints} can be written as a sum of terms, such that the BFKL eigenvalue in each channel only depends on the number of crossing singularities of the type $\widetilde{\Yf}'_{1/3,s}=-1$ for the index $s$ corresponding to that channel.
Furthermore, using target-projectile symmetry, we have shown that all BFKL eigenvalues contributing to any $n$-point remainder function in the multi-Regge limit are multiples of the two-Reggeon BFKL eigenvalue,
			\begin{equation}
				e_n = k\cdot\left(\log(1+\sqrt{2})-\sqrt{2}\right) = k\cdot e_2,
				\nonumber
			\end{equation}
where $k\in \mathbb{N}$.
The corresponding crossing pattern is also determined uniquely and is described by $n_{1/3, s_0}^{(+)}=n_{1/3, s_0}^{(-)}=k$ for the channel $s_0$ in which the corresponding BFKL eigenvalue appears.
		\end{minipage}
	};
\end{tikzpicture}
\end{center}
Considering the picture of $m$-Reggeon bound states suggested by the Wilson loop OPE (see section \ref{sec:intro}), it is most natural to expect that the BFKL eigenvalue of the $m$-Reggeon bound state at strong coupling is given by $(m-1)$ times the two-Reggeon BFKL eigenvalue, $e_n\stackrel{?}{=}(n-1)\, e_2$.
However, based on the method used in this paper, we cannot prove (or disprove) this expectation as this would require some ``dynamic'' information on the analytic continuation, e.g. which cross ratios were analytically continued, to characterize the specific Mandelstam regions in which the $m$-Reggeon bound states are expected to appear.
\section{Example: the nine-point remainder function}
\label{sec:three_reggeon_bs}

To fill the general results of the preceding sections with life, we discuss the concepts introduced above for the example of the nine-point amplitude with a focus on the three-Reggeon bound state, which has not yet been investigated in the strong coupling limit and whose BFKL eigenvalue is still unknown. 
We start by introducing the relevant formulas for the nine-point remainder function before briefly discussing the crossing patterns which are expected to describe the two-Reggeon bound state and show that the resulting remainder function is indeed characterized by the two-Reggeon BFKL eigenvalue $e_2$ already obtained in the six-point case.
Furthermore, we show that such a solution can be found in all channels in which the two-Reggeon bound state is expected to appear by Regge theory. 
Finally we analyze the most natural candidate of crossing patterns for the three-Reggeon bound state (see the discussion at the end of section \ref{sec:rem_fct}) and study the corresponding Bethe ansatz equations and the BFKL eigenvalue their solution leads to.
Additionally, we show that this solution can be found in all channels in which the three-Reggeon bound state is expected to be found in the nine-point case and show that the simple structure of this solution being a multiple of the two-Reggeon BFKL eigenvalue also holds for the kinematically subleading terms studied in \cite{Sprenger:2016jtx}.\par
Recall from the discussion in section \ref{sec:rem_fct} that the three-Reggeon bound state is expected to appear for eight gluons or more.
We choose to study the nine-point amplitude as this allows us to check whether the crossing pattern conjectured to describe the three-Reggeon bound state leads to the same BFKL eigenvalue in all channels in which the three-Reggeon bound state is expected to appear from Regge theory.
Furthermore, choosing the nine-point amplitude avoids the technical difficulties introduced by the $A_{\mathrm{extra}}$-contribution present in the case $n=4k$, which we alluded to before.

\subsection{The nine-point amplitude and Mandelstam regions}
\label{sec:9pointamplitude}
Let us start by discussing the kinematics of the nine-point case.
We can construct 18 dual conformal cross ratios, four triplets of independent cross ratios, as well as six additional ones which are related to the independent cross ratios through conformal Gram relations.
In the MRL, the large cross ratios $u_{1,s}\rightarrow 1$ and the small ones $u_{2/3,s} \rightarrow 0$, while the dependent cross ratios go to one, as well.\par
The different contributions to the remainder function were described in general in section \ref{sec:review}, and their explicit form for $n=9$ reads
\begin{align} 
	A_{\text{free}}=&\sum_{s=1}^4 \frac{|m_s|}{2 \pi} \int\displaylimits_\mathbb{R} d\theta\ \cosh\theta\ \text{log}\left((1+\widetilde{\Yf}_{1,s})(1+\widetilde{\Yf}_{2,s})^{\sqrt{2}}(1+\widetilde{\Yf}_{3,s})\right),\label{eq:Afree9} \\
A_{\text{per}}=&-\frac{1}{2} \left(|m_2|^2+|m_3|^2+m_1 \bar{m}_3+m_3\bar{m}_1+m_2\bar{m}_4+m_4\bar{m}_2\right)\nonumber \\
&-\frac{\sqrt{2}}{4}\left(m_1\bar{m}_2+m_2\bar{m}_1+m_1\bar{m}_4+m_4\bar{m}_1\right.\nonumber \\
&\quad\quad\quad\quad\left.+m_3\bar{m}_4+m_4\bar{m}_3+2m_2\bar{m}_3+2m_3\bar{m}_2\right). \label{eq:Aper9} 
\end{align}
Following the discussion in section \ref{sec:rem_fct}, we refrain from spelling out the contribution $\Delta$ as it is a rather lengthy expression, which only contributes phases to the remainder function in the Mandelstam regions and is therefore not relevant in the following (the contribution can be found, for example, in \cite{Yang:2010as}).\par
As mentioned before, the remainder function in the Euclidean region, where all energy variables are negative, is trivial.
This behavior is obvious for the $A_\text{free}$-contribution since in the multi-Regge limit all integrals are negligible.
However, the behavior is less obvious for the other contributions, but as it turns out the remaining contributions, $A_\text{per}$ and $\Delta$, cancel each other in the multi-Regge limit.
Thus, the BDS ansatz completely describes the scattering amplitude in the Euclidean region, which agrees with the field theory predictions (see \cite{Bartels:2008ce}).
Therefore, we proceed by looking at Mandelstam regions, in which the remainder function does not vanish.
\subsubsection*{Two-Reggeon bound states}
Based on the analysis presented in sections \ref{sec:constraints} and \ref{sec:rem_fct}, we can parametrize the allowed BFKL eigenvalues by the number of crossing singularities $n_{1/3, s}^{(+/-)}$.
Given that the BFKL eigenvalue of the two-Reggeon bound state is realized by a pair of crossing solutions $n_{1/3, s}^{(+/-)}=1$ in a particular channel in the six- and seven-point remainder function (see \cite{Bartels:2010ej, Bartels:2013dja, Bartels:2014ppa, Bartels:2014mka, Sprenger:2016jtx}), we proceed by analyzing this configuration in the nine-point case to check whether our calculations reproduce the expected results for the two-Reggeon bound state regions.\par
In the nine-point case there are many kinematic regions in which the two-Reggeon bound state is expected to appear, the simplest regions describing so-called short Regge cuts, for which the two-Reggeon bound states only propagate in a single $t$-channel.
As a consequence, the remainder function is expected to depend on a single $s$-like Mandelstam variable $s_i$.
In terms of the parameters (\ref{eq:def_mrl_param}), this corresponds to a dependence on a single $\varepsilon_s$ or, equivalently, a single triplet of cross ratios.
These cuts are connected to a kinematic region in which two of the produced particles are chosen to be incoming.
As mentioned above, we expect these regions to be realized by one pair of crossing singularities of the type $\Yt'_{1/3,s}(\theta)=-1$ with endpoints in $\pm\frac{i\pi}{4}$.\par
\begin{figure}[t]
\centering
  \includegraphics[width=1\textwidth]{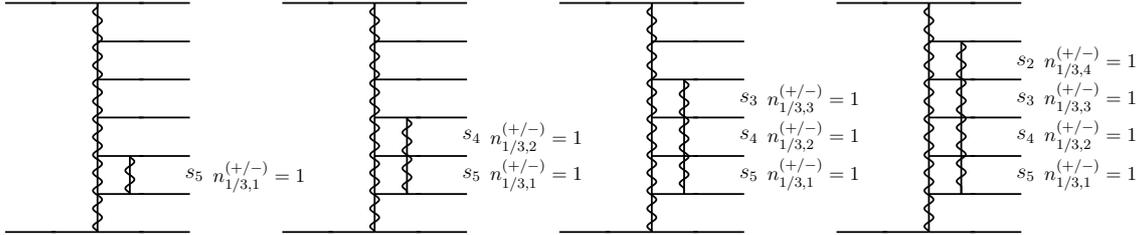}
\caption{Two-Reggeon bound states spanning different channels and their realization given in terms of the numbers $n^{(+/-)}_{1/3,s}$ of crossing singularities in the corresponding channels.}
\label{fig:2Reggeonboundstates}
\end{figure}

For $n>6$, the Regge cut induced by the two-Reggeon bound state can span several $t$-channels, in which case the remainder function depends on several triplets of cross ratios.
These cuts are usually referred to as long cuts and we expect those to be realized in the nine-point case by a pair of crossing singularities of the $\Yt_{1/3,s}$-functions associated to the corresponding $t$-channels. In figure~\ref{fig:2Reggeonboundstates} we illustrate two-Reggeon bound states of different lengths corresponding to the regions described in table~\ref{tab:tworeggeonbs}.
Analyzing these configurations of crossing singularities, we find that they are all described by the two-Reggeon BFKL eigenvalue $e_2=\log \left( 1+\sqrt{2}\right)-\sqrt{2}$, as expected, and that they have the kinematic dependence expected from Regge theory.
We spell out the analyzed remainder functions and the kinematic regions, which we expect them to describe, in table~\ref{tab:tworeggeonbs}.
\begin{table}[htb]
\begin{adjustbox}{center}
\renewcommand{\arraystretch}{1.4}
\begin{tabular}{|l|l|}
	\hline {\# of crossing singularities $n_{1/3,s}^{(+/-)}$} & {Remainder function}\\  \hline  
	$n_{1/3, 1}^{(+)}=n_{1/3, 1}^{(-)}=1$ & $\left.e^{R_{9,+++--}}\right|_{\mathrm{MRL}}\sim(1-u_{1,4})^{\frac{\sqrt \lambda}{2\pi}\,e_2}$\\ \hline
	$n_{1/3, 1}^{(+)}=n_{1/3, 1}^{(-)}=n_{1/3, 2}^{(+)}=n_{1/3, 2}^{(-)}=1$  & $\left.e^{R_{9,++---}}\right|_{\mathrm{MRL}}\sim\left( (1-u_{1,3})(1-u_{1,4})\right)^{\frac{\sqrt \lambda}{2\pi}\,e_2}$\\ \hline
	$\!\begin{aligned}[t]
	&n_{1/3, 1}^{(+)} =n_{1/3, 1}^{(-)}=n_{1/3, 2}^{(+)}=n_{1/3, 2}^{(-)}=\\
	&n_{1/3, 3}^{(+)}=n_{1/3, 3}^{(-)}=1\end{aligned}$ & $\left.e^{R_{9, +----}}\right|_{\mathrm{MRL}}\sim\left( (1-u_{1,2})(1-u_{1,3})(1-u_{1,4})\right)^{\frac{\sqrt \lambda}{2\pi}\,e_2}$\\ \hline
$\!\begin{aligned}[t]
	&n_{1/3, 1}^{(+)} =n_{1/3, 1}^{(-)}=n_{1/3, 2}^{(+)}=n_{1/3, 2}^{(-)}=\\
	&n_{1/3, 3}^{(+)}=n_{1/3, 3}^{(-)}=n_{1/3, 4}^{(+)}=n_{1/3, 4}^{(-)}=1\end{aligned}$ & $\left.e^{R_{9,-----}}\right|_{\mathrm{MRL}}\sim\left( (1-u_{1,1})(1-u_{1,2})(1-u_{1,3})(1-u_{1,4})\right)^{\frac{\sqrt \lambda}{2\pi}\,e_2}$ \\ \hline
\end{tabular}
\renewcommand{\arraystretch}{1}
\end{adjustbox}
\caption{The different types of Regge cuts due to the two-Reggeon bound states all give rise to the same BFKL eigenvalue $e_2=\log (1+\sqrt{2})-\sqrt{2}$ in the remainder function, which agrees with the value found in the six-point case.}\label{tab:tworeggeonbs}
\end{table}

\subsection{A solution with BFKL eigenvalue $e_3=2\cdot e_2$}
\label{sec:BFKL2e2}

\begin{figure}[b]
\centering
\begin{minipage}{.5\textwidth}
  \centering
  \includegraphics[width=0.75\textwidth]{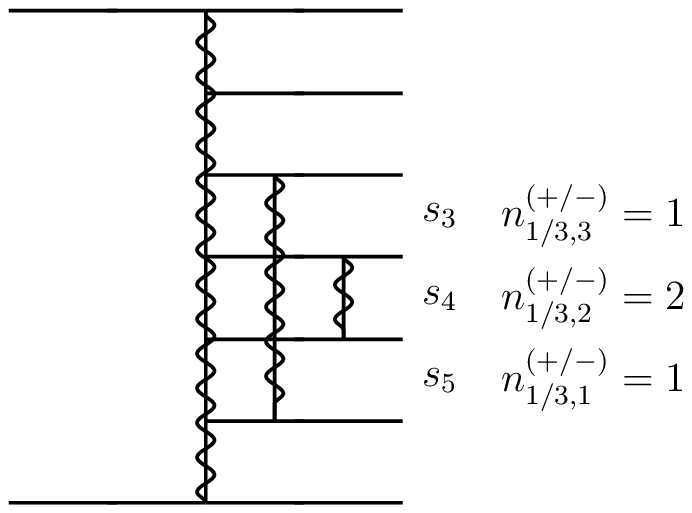}
\end{minipage}%
\begin{minipage}{.5\textwidth}
  \centering
  \includegraphics[width=0.75\textwidth]{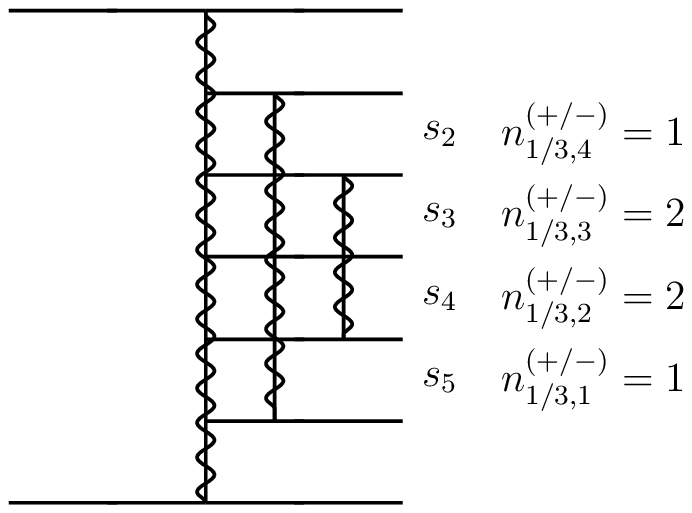}
\end{minipage}
\caption{Left: The three-Reggeon bound state is described by an additional Reggeon in the corresponding $t$-channel. We study the three-Reggeon bound state connected to the $t_4$-channel. Such a bound state can only appear if the number of Reggeons in neighboring channels differs by at most one, which is satisfied by the shown configuration of crossing singularities. The figure also shows the number $n^{(+/-)}_{1/3,s}$ of crossing singularities which occur in the corresponding channels. Right: A configuration in which the three-Reggeon bound state appears in both the $t_3$- and $t_4$-channel. Again the associated crossing singularities are illustrated.}
\label{fig:3Reggeonboundstate}
\end{figure}

\begin{figure}
\centering
\includegraphics[width=\textwidth]{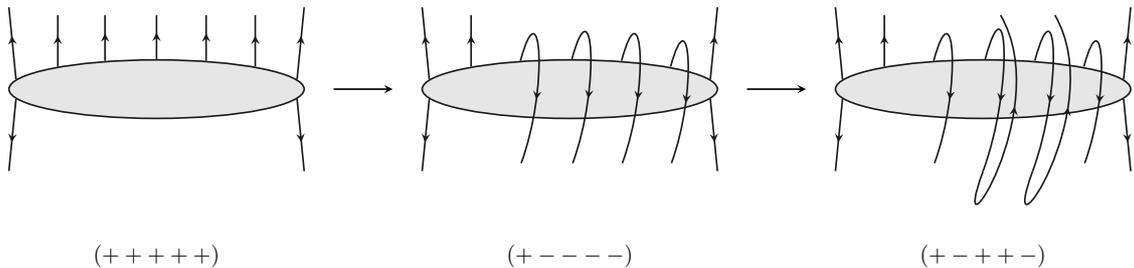}
\caption{Graphical representation of the analytic continuation to the Mandelstam region in which the three-Reggeon bound state is expected to contribute to the nine-point remainder function. During the first analytic continuation the particles $p_5,\dots, p_8$ are chosen to be incoming and cross the center of the diagram which corresponds to the long Regge cut of the region $(+----)$, which is described by two-Reggeon bound states only. To obtain a bound state of more Reggeons in the $t_4$-channel, a second analytic continuation needs to be performed choosing the particles $p_6$ and $p_7$ to be outgoing again, during which we expect two additional crossing solutions in the $\Yt_{a,2}$-functions.}
\label{fig:3Rcutregion}
\end{figure}

In this section we now investigate the three-Reggeon bound state and its BFKL eigenvalue $e_3$.
In the nine-point case the new Reggeon bound state can appear either in the $t_3$- or the $t_4$-channel (or both) and we choose to discuss the region with the three-Reggeon bound state in the $t_4$-channel which is illustrated in figure \ref{fig:3Reggeonboundstate}.
The analysis for the $t_3$-channel is analogous and leads to the same results.
As we have explained at the end of section \ref{sec:rem_fct}, the BFKL eigenvalue of the three-Reggeon bound state cannot be fixed uniquely by analyzing the Bethe ansatz equations at the endpoint alone.
Therefore, in this section, we analyze a crossing pattern for which the most natural conjecture for the BFKL eigenvalue of the three-Reggeon bound state, namely that $e_3=2e_2$, is realized.
While it is still a conjecture that this configuration describes the three-Reggeon bound state, this configuration is consistent with the expectation from integrability that the BFKL eigenvalue of the three-Reggeon bound state is given by a sum of two BFKL eigenvalues of two-Reggeon bound states. 
Furthermore, to reach the Mandelstam region in which the three-Reggeon bound state is expected to appear, two subsequent analytic continuations of the cross ratios (as illustrated in figure \ref{fig:3Rcutregion}) are necessary.
More specifically, in the first analytic continuation the four external particles $p_5,\dots, p_8$ are chosen to be incoming, leading to the Mandelstam region $(+----)$.
By analogy with the six- and seven-point cases, we expect this region to be described by a pair of crossing singularities of the type $\widetilde{\Yf}'_{a,s}=-1$ for the indices $s=1,2,3$, as those lead to contributions in the $t_3$-, $t_4$- and $t_5$-channel, respectively (see eq.\ (\ref{eq:rem_fct_endp})).
It is thus natural to expect that in the second analytic continuation, in which the Mandelstam region $(+-++-)$ is reached by choosing particles $p_6$, $p_7$ to be outgoing again (see figure \ref{fig:3Rcutregion}), two further crossing singularities occur in the $\Yt_{a,2}$-function connected to the $t_4$-channel.
Based on this expectation, we study the crossing pattern
\begin{align}
	&n_{1/3, 3}^{(+)} = n_{1/3,3}^{(-)}=1,\nonumber\\
	&n_{1/3, 2}^{(+)} = n_{1/3,2}^{(-)}=2,\nonumber\\
	&n_{1/3, 1}^{(+)} = n_{1/3,1}^{(-)}=1.\label{eq:cross_example}
\end{align}
Note that we include crossing singularities not only for the $t_4$-channel, in which we expect the three-Reggeon bound state to appear, but also in the adjacent channels.
The reason for this is that in the planar limit, the number of Reggeons propagating in a given channel may only differ by one from the number of Reggeons propagating in the adjacent channels (see, for example, \cite{Lipatov:2009nt, Bartels:2011nz, Bargheer:2016eyp}).
Thus, two-Reggeon bound states need to appear in both the $t_3$- and $t_5$-channel, which are realized (as in the six- and seven-point case) as one pair of crossing solutions as indicated in eq.\ (\ref{eq:cross_example}).
\par
Having fixed the crossing pattern, we can immediately follow the general analysis presented in section \ref{sec:rem_fct} to calculate the nine-point remainder function for this region, since the endpoints of the crossing singularities are fixed by the general analysis of section \ref{sec:constraints}.
In this case, the parameters $\varepsilon_s'$ and $w_s'$ at the endpoint read
\begin{alignat}{2}
\varepsilon_1'&=\varepsilon_1,\qquad &w_1'&=\gamma\,w_1,\nonumber\\
\varepsilon_2'&=\frac{1}{\gamma}\,\varepsilon_2,\qquad  &w_2'&=w_2,\nonumber\\
\varepsilon_3'&=\varepsilon_3,\qquad  &w_3'&=\frac{1}{\gamma}w_3,\nonumber\\
\varepsilon_4'&=\sqrt{\gamma}\varepsilon_4,\qquad &w_4'&=\frac{1}{\sqrt{\gamma}}\,w_4,
\label{eq:epsilonomegatriplebs}
\end{alignat}
where $\gamma =-\,(1+\sqrt{2}\,)^2$, see eq.\ (\ref{eq:rel_aper}).
Using these relations, we can calculate the contributions to the remainder function at the endpoint of the continuation, $A_\text{free}'$ and $A_\text{per}'$, and find
\begin{align}
A_\text{free}'\cong\, \sqrt{2}\log \varepsilon_1+2\,\sqrt{2}\log \varepsilon_2+\sqrt{2}\log \varepsilon_3+\dots,
\end{align}
see eq.\ (\ref{eq:afree_endp_rem}) and
\begin{equation}
	A_\text{per}'-A_\text{per}\cong-\log\left(1+\sqrt{2}\right) \log \varepsilon_1 - 2\log\left(1+\sqrt{2}\right)  \log\varepsilon_2-\log\left(1+\sqrt{2}\right) \log \varepsilon_3+\dots,
\end{equation}
see eq.\ (\ref{eq:delta_aper_mt}), where the dots indicate constants and phases which cannot be fixed by our approach as explained in section \ref{sec:rem_fct}, but which do not affect the BFKL eigenvalue.
Collecting the contributions to the remainder function we find
\begin{equation}
A_\text{per}'-A_\text{per}+A_\text{free}'\cong -(2\cdot e_2) \log \varepsilon_2 -e_2 \log (\varepsilon_1\cdot \varepsilon_3)+\dots,
\end{equation}
and the structure of the remainder function is then given by
\begin{equation}
	\left.e^{R_{9,+-++-}}\right|_{\text{MRL}} \sim (1-u_{1,3})^{\frac{\sqrt{\lambda}}{2\pi} (2\cdot e_2)}\cdot\left((1-u_{1,2})\cdot (1-u_{1,4})\right)^{\frac{\sqrt{\lambda}}{2\pi} e_2},
\end{equation}
with the two-Reggeon BFKL eigenvalue $e_2=\log (1+\sqrt{2})-\sqrt{2}$, which is the same as in the six-point case. Note that the three-Reggeon BFKL eigenvalue $e_3=2\cdot e_2$ is indeed the sum of two BFKL eigenvalues of two-Reggeon bound states, as claimed before.
In the next section, we analyze two properties of this crossing pattern, namely whether the same solution can be realized in other channels as required by Regge theory, and whether this simple structure for the BFKL eigenvalue of the three-Reggeon bound state also holds for contributions to the remainder function, which are kinematically subleading in the multi-Regge limit.
\subsubsection{Consistency of the crossing pattern and subleading kinematics}
\label{sec:consistency}
In this section, we show that the crossing pattern under consideration passes several consistency checks.
First, consistency of our results with Regge theory requires that, in different Mandelstam regions, the three-Regge bound state can appear in either the $t_3$- or $t_4$-channel and additionally as a long triple-Regge cut simultaneously in the $t_3$- and $t_4$-channel as illustrated in figure \ref{fig:3Reggeonboundstate}.
In all these cases, the three-Reggeon bound state needs to appear with the same BFKL eigenvalue.
The three Mandelstam regions, in which we expect the three-Reggeon bound state to appear, are the regions $(+-++-)$, $(-++-+)$ and $(-+++-)$.
The first of these regions was analyzed in the previous section.
Performing the corresponding calculations for the crossing patterns
\begin{align*}
	&n_{1/3, 4}^{(+)} = n_{1/3,4}^{(-)}=1,\\
	&n_{1/3, 3}^{(+)} = n_{1/3,3}^{(-)}=2,\\
	&n_{1/3, 2}^{(+)} = n_{1/3,2}^{(-)}=1
\end{align*}
and 
\begin{align*}
	&n_{1/3, 4}^{(+)} = n_{1/3,4}^{(-)}=1,\\
	&n_{1/3, 3}^{(+)} = n_{1/3,3}^{(-)}=2,\\
	&n_{1/3, 2}^{(+)} = n_{1/3,2}^{(-)}=2,\\
	&n_{1/3, 1}^{(+)} = n_{1/3,1}^{(-)}=1,
\end{align*}
we find the remainder functions
\begin{align}
	&\left.e^{R_{9,+-++-}}\right|_{\text{MRL}} \sim (1-u_{1,3})^{\frac{\sqrt{\lambda}}{2\pi} (2\cdot e_2)}\cdot\left( (1-u_{1,2})\cdot (1-u_{1,4}) \right)^{\frac{\sqrt{\lambda}}{2\pi} e_2},\nonumber \\
	&\left.e^{R_{9,-++-+}}\right|_{\text{MRL}} \sim (1-u_{1,2})^{\frac{\sqrt{\lambda}}{2\pi} (2\cdot e_2)}\cdot\left( (1-u_{1,1})\cdot (1-u_{1,3})\right)^{\frac{\sqrt{\lambda}}{2\pi} e_2},\nonumber \\
	&\left.e^{R_{9,-+++-}}\right|_{\text{MRL}} \sim \left( (1-u_{1,2})\cdot (1-u_{1,3}) \right)^{\frac{\sqrt{\lambda}}{2\pi} (2\cdot e_2)}\cdot\left( (1-u_{1,1})\cdot (1-u_{1,4})\right)^{\frac{\sqrt{\lambda}}{2\pi} e_2}.
\end{align}
Indeed, these solutions all have the same two- and three-BFKL eigenvalues $e_3=2\cdot e_2$ and $e_2$ in the relevant $t$-channels.
Furthermore, note that the remainder functions are consistent with target-projectile symmetry (see eq.\ (\ref{eq:tp_sym_crs})), since the first two regions are related by target-projectile symmetry on the level of the cross ratios and the remainder function while the third region is invariant under the symmetry.\par
Lastly, we analyze whether the contributions from the three-Reggeon bound state to the remainder function which are subleading in kinematics also have the simple structure found for the leading term.
More specifically, as shown for the six- and seven-point amplitude in \cite{Sprenger:2016jtx}, the remainder function can be expanded in a series of terms of the form $\sim\frac{\log^k w}{\log^{n}\varepsilon}$, which vanish in the strict multi-Regge limit and are kinematically subleading compared to terms $\sim\log\varepsilon$ considered so far (see eq.\ (\ref{eq:rem_fct_endp})). 
While subleading, these terms are interesting as they contain additional information on the BFKL eigenvalue and allow a precise comparison of the TBA at strong coupling with the finite-coupling expression of the six-point remainder function in the multi-Regge limit \cite{Basso:2014pla}.
These subleading terms arise as corrections to the relations between the kinematical parameters at the endpoint and the starting point of the continuation (see eqs.\ (\ref{eq:crs_signs}) and (\ref{eq:rel_aper})) and as higher order terms appearing in the contributions $A_{\mathrm{free}}$ and $A_{\mathrm{per}}$ to the remainder function (see eqs.\ (\ref{eq:afree_endp_rem}) and (\ref{eq:delta_aper_mt})).
Since the analysis is identical to the six- and seven-point cases, we refer the reader to \cite{Sprenger:2016jtx} for technical details.
For the purpose of this analysis, we again study the region $(+-++-)$ described by the crossing pattern (\ref{eq:cross_example}).
To find the relevant contributions at each order in the kinematic expansion $\sim\frac{1}{\log^n\varepsilon}$ we start from a general ansatz for the kinematic variables at the endpoint of the continuation. 
For example, the ansatz for the first subleading order reads
\begin{align}
\varepsilon_s'&=\varepsilon_s\left(a_{s,1}+\frac{a_{s,2}}{\log\varepsilon_{s-1}}+\frac{a_{s,3}}{\log\varepsilon_s}+\frac{a_{s,4}}{\log\varepsilon_{s+1}}+\mathcal{O}(\log^{-2}\varepsilon)\right),\nonumber\\
w_s'&=w_s\left(b_{s,1}+\frac{b_{s,2}}{\log\varepsilon_{s-1}}+\frac{b_{s,3}}{\log\varepsilon_s}+\frac{b_{s,4}}{\log\varepsilon_{s+1}}+\mathcal{O}(\log^{-2}\varepsilon)\right).
\label{eq:ansatz_subleading}
\end{align}
We then solve the condition $u_{a,s}'=u_{a,s}$ order-by-order in $1/\log\varepsilon$ to fix the coefficients $a_{s,i}$ and $b_{s,i}$.
Note that at subleading order the kinematic variables from one channel at the endpoints of the continuation depend on the variables from neighboring channels, as well, which is not the case at leading order. 
The resulting parameters to the first subleading order are presented in appendix \ref{app:subleading}.\par
We then extend the ansatz for $\varepsilon_s'$ and $w_s'$ (\ref{eq:ansatz_subleading}) by including higher-order terms in $1/\log\varepsilon$, and solve the equations $u_{a,s}'=u_{a,s}$ to fix the parameters in the ansatz order-by-order to any desired order in the subleading terms.
Once the kinematic parameters at the endpoints of the continuation are fixed up to the desired order in $1/\log\varepsilon$ we can compute the subleading terms in the remainder function.
Up to third order in the subleading terms, we find the following result for the remainder function for the crossing pattern (\ref{eq:cross_example}):
\begin{align}
\left.R_{9,+-++-}\right|_{\mathrm{MRL}}=-\frac{\sqrt{\lambda}}{2\pi}\,&\left[(\sqrt{2}-\log(1+\sqrt{2}\,))\,\log\varepsilon_1-\sqrt{2}\,\frac{\log^2 w_1}{\log\varepsilon_1}-\frac{5}{6\sqrt{2}}\frac{\log^4 w_1}{\log^3\varepsilon_1}\right.\nonumber\\
&+ 2\,(\sqrt{2}-\log(1+\sqrt{2}\,))\,\log\varepsilon_2-2\,\sqrt{2}\,\frac{\log^2 w_2}{\log\varepsilon_2}-2\,\frac{5}{6\sqrt{2}}\frac{\log^4 w_2}{\log^3\varepsilon_2}\nonumber\\
&+\left.(\sqrt{2}-\log(1+\sqrt{2}\,))\,\log\varepsilon_3-\sqrt{2}\,\frac{\log^2 w_3}{\log\varepsilon_3}-\frac{5}{6\sqrt{2}}\frac{\log^4 w_3}{\log^3\varepsilon_3}\,\right]\nonumber\\
&\quad+\mathcal{O}\left(\log^{-4}\varepsilon\right)\label{eq:subleadingShort},
\end{align}
where we have only spelled out the contributions with the maximal power of $\log w$ for each order of $1/\log\varepsilon$, namely the terms of the form $\sim\frac{\log^{k+1} w}{\log^{k}\varepsilon}$.
Note that the subleading contributions in eq.\ (\ref{eq:subleadingShort}) in the channel associated with the three-Reggeon bound state again are simply given as twice the corresponding contributions for the two-Reggeon bound states, as we established before for the leading terms.\par
This is a very interesting result for the following reason: recall that at weak and finite coupling, remainder functions in the multi-Regge limit are naturally described in terms of certain Fourier-Mellin integrals (see, for example, \cite{Basso:2014pla, DelDuca:2016lad, DelDuca:2019tur}). 
At strong coupling, these Fourier-Mellin integrals can be evaluated by a saddle point approximation, which then reproduces the results from the strong coupling TBA discussed in this paper (see \cite{Basso:2014pla, Sprenger:2016jtx}).
In particular, the BFKL eigenvalue of a $m$-Reggeon bound state $\omega_m(\{\nu_i, n_i\})$ ($i=1,\dots,m-1$) appearing in a given Fourier-Mellin integral would then become the BFKL eigenvalues discussed in this paper, $\omega_m(\{\nu_i, n_i\}) \rightarrow e_m$.
The subleading terms discussed above arise from the Fourier-Mellin integrals when choosing the terms $\frac{\log^{k}w}{\log^{n}\varepsilon}$ to be small but finite, because the saddle point moves slightly away from the point where those terms strictly vanish (see \cite{Sprenger:2016jtx}).
Importantly, the equation determining the location of the saddle point depends on the full structure of the BFKL eigenvalue $\omega_m(\{\nu_i, n_i\})$ and not just its value on the saddle point.
In particular, as shown in \cite{Sprenger:2016jtx}, the terms of the form $\sim\frac{\log^{k+1}w}{\log^{k}\varepsilon}$ shown in eq.\ (\ref{eq:subleadingShort}) do not depend on the impact factors and thus only probe the structure of the BFKL eigenvalues appearing in the Fourier-Mellin integrals.
Hence, assuming that a similar Fourier-Mellin integral describing the region under consideration here exists, our result (\ref{eq:subleadingShort}) provides evidence that the BFKL eigenvalue of the three-Reggeon bound state is given by a sum of two two-Reggeon BFKL eigenvalues, not just on the strong coupling saddle point, but already for the BFKL eigenvalue appearing in the Fourier-Mellin integral.
\section{Conclusions}
\label{sec:conclusions}
In this paper, we have studied the constraints the multi-Regge limit imposes on the $n$-point remainder function in different kinematic regimes, which are reached by analytic continuations of the $\Yf$-system describing scattering amplitudes at strong coupling.
During these analytic continuations, singularities of the $\Yf$-system may cross the integration contour and lead to a non-trivial remainder function at the endpoint of the continuation.
We have shown that the possible patterns of crossing singularities are very simple and can be parametrized easily.
Based on this result, we have shown that all possible BFKL eigenvalues are multiples of the two-Reggeon BFKL eigenvalue.
Furthermore, based on the example of the nine-point amplitude, we have provided evidence that this additivity of BFKL eigenvalues is also satisfied for kinematically subleading terms, suggesting that this simple structure of BFKL eigenvalues also holds beyond the saddle point governing the strong coupling solution.\par
This paper answers the questions \textbf{which} BFKL eigenvalues are possible in any analytic continuation consistent with the multi-Regge limit.
However, it does not answer the questions \textbf{if} and \textbf{when} these eigenvalues appear, i.e.\ in which kinematic region they contribute and to which Reggeon bound state they are associated.
In fact, it is not even clear that bound states of more than two Reggeons are visible at strong coupling.
In the six-point case, the strong coupling result obtained from the TBA arises from the finite-coupling Fourier-Mellin integral as a saddle point of that integral.
However, as shown in \cite{Bartels:2013jna, Bartels:2014jya, Bartels:2020twc}, for more than six gluons the remainder function in certain kinematic regions is described by a sum of several such integrals.
At strong coupling, the remainder function obtained from the TBA should thus represent the dominant saddle point of those terms.
Whether this dominant saddle point necessarily contains the interesting bound states of more than two Reggeons, however, is not a priori clear.
So far, this question can only be answered by performing the analytic continuation numerically, which becomes more and more difficult as the number of gluons is increased.
Whether there is a way along the approach proposed in this paper to predict which BFKL eigenvalue appears in which kinematic region is a very interesting open question, which we leave for future research.
\acknowledgments
We would like to thank Benjamin Basso, Simon Caron-Huot and Volker Schomerus for many helpful discussions.
The work of TA is partially supported by a Durham Doctoral Studentship.
The work of MS was partially supported by the Swiss National Science Foundation through the NCCR SwissMAP.
\appendix

\section{$\Yf$-system kernels and S-matrices}
\subsection{Integration kernels of the $\Yf$-system}
\label{app:kernels}
To state the kernels of the $\Yf$-system (\ref{eq:def_tba}), we start from the case $\varphi_s=0$ (note that in this case the complex parameters $m_s=|m_s|e^{i\varphi_s}$ are real and that $\widetilde{\Yf}_{a,s}=\Yf_{a,s}$), for which the $\Yf$-system equations are given by (see \cite{Alday:2010vh})
\begin{align}
	&\log\Yf_{1,s}=-m_s\cosh\theta-C_s-\frac{1}{2}K_2\star\beta_s-K_1\star\alpha_s-\frac{1}{2}K_3\star\gamma_s,\nonumber\\
	&\log\Yf_{2,s}=-\sqrt{2}m_s\cosh\theta-K_2\star\alpha_s-K_1\star\beta_s,\nonumber\\
	&\log\Yf_{3,s}=-m_s\cosh\theta+C_s-\frac{1}{2}K_2\star\beta_s-K_1\star\alpha_s+\frac{1}{2}K_3\star\gamma_s,
	\label{eq:ysys_app}
\end{align}
where the convolution is defined as
\begin{equation}
	(K_a\star f)(\theta)=\int\limits_{\mathbb{R}}d\theta'K_a(\theta-\theta')f(\theta'),
	\label{eq:def_conv}
\end{equation}
and the three basic kernels are given by
\begin{equation}
	K_1(\theta)=\frac{1}{2\pi}\frac{1}{\cosh\theta}, \quad K_2(\theta)=\frac{\sqrt{2}}{\pi}\frac{\cosh\theta}{\cosh 2\theta}, \quad K_3(\theta)=\frac{i}{\pi}\tanh 2\theta.
	\label{eq:basic_kernels}
\end{equation}
Furthermore, certain combinations of $\Yf$-functions $\alpha_s$, $\beta_s$ and $\gamma_s$ are defined as
\begin{align}
	&\alpha_s=\log\frac{(1+\Yf_{1,s})(1+\Yf_{3,s})}{(1+\Yf_{2,s-1})(1+\Yf_{2,s+1})},\nonumber\\
	&\beta_s=\log\frac{(1+\Yf_{2,s})^2}{(1+\Yf_{1,s-1})(1+\Yf_{1,s+1})(1+\Yf_{3,s-1})(1+\Yf_{3,s+1})},\nonumber\\
	&\gamma_s=\log\frac{(1+\Yf_{1,s-1})(1+\Yf_{3,s+1})}{(1+\Yf_{1,s+1})(1+\Yf_{3,s-1})},
\end{align}
from which an arbitrary kernel between two $\Yf$-functions $\mathcal{K}^{a,a'}_{s,s'}$ can be easily read off.
For the case relevant in the main text, namely that of non-vanishing parameters $\varphi_s$, the $\Yf$-system (\ref{eq:def_tba}) is obtained by simply making the replacements
\begin{equation}
	m_s\rightarrow |m_s|,\quad \Yf_{a,s}\rightarrow \widetilde{\Yf}_{a,s}, \quad \mathcal{K}_{s,s'}^{a,a'}(\theta-\theta')\rightarrow\mathcal{K}_{s,s'}^{a,a'}(\theta-\theta'+i\varphi_s-i\varphi_{s'}),
	\label{eq:ysys_repl}
\end{equation}
where $\widetilde{\Yf}_{a,s}(\theta):=\Yf_{a,s}(\theta+i\varphi_s)$, as in the main text.
Explicitly, the kernels of the $\Yf$-system (\ref{eq:def_tba}) thus read:
\begin{alignat}{2}
&\mathcal{K}_{s,s}^{1,2\pm1}&&=-K_1(\theta-\theta'), \nonumber\\
&\mathcal{K}_{s,s}^{1,2}&&=-K_2(\theta-\theta'), \nonumber\\
&\mathcal{K}_{s,s}^{2,2\pm 1}&&=-K_2(\theta-\theta'), \nonumber\\
&\mathcal{K}_{s,s}^{2,2}&&=-2 K_1(\theta-\theta'), \nonumber\\
&\mathcal{K}_{s,s}^{3,2\pm 1}&&=-K_1(\theta-\theta'), \nonumber\\
&\mathcal{K}_{s,s}^{3,2}&&=-K_2(\theta-\theta'), \nonumber\\
&\mathcal{K}_{s,s\pm 1}^{1,2}&&=K_1(\theta-\theta'+i\varphi_s-i\varphi_{s\pm 1}), \nonumber\\
&\mathcal{K}_{s,s\pm 1}^{1,1}&&=\frac{1}{2} K_2(\theta-\theta'+i\varphi_s-i\varphi_{s\pm 1})\pm \frac{1}{2} K_3(\theta-\theta'+i\varphi_s-i\varphi_{s\pm 1}),\nonumber\\ 
&\mathcal{K}_{s,s\pm 1}^{1,3}&&=\frac{1}{2} K_2(\theta-\theta'+i\varphi_s-i\varphi_{s\pm 1})\mp \frac{1}{2} K_3(\theta-\theta'+i\varphi_s-i\varphi_{s\pm 1}),\nonumber\\ 
&\mathcal{K}_{s,s\pm 1}^{2,2}&&=K_2(\theta-\theta'+i\varphi_s-i\varphi_{s\pm 1}),\nonumber\\
&\mathcal{K}_{s,s\pm 1}^{2,2\pm 1}&&=K_1(\theta-\theta'+i\varphi_s-i\varphi_{s\pm 1}),\nonumber\\
&\mathcal{K}_{s,s\pm 1}^{3,2}&&=K_1(\theta-\theta'+i\varphi_s-i\varphi_{s\pm 1}),\nonumber\\
&\mathcal{K}_{s,s\pm 1}^{3,1}&&=\frac{1}{2} K_2(\theta-\theta'+i\varphi_s-i\varphi_{s\pm 1})\mp \frac{1}{2} K_3(\theta-\theta'+i\varphi_s-i\varphi_{s\pm 1}),\nonumber\\ 
&\mathcal{K}_{s,s\pm 1}^{3,3}&&=\frac{1}{2} K_2(\theta-\theta'+i\varphi_s-i\varphi_{s\pm 1})\pm \frac{1}{2} K_3(\theta-\theta'+i\varphi_s-i\varphi_{s\pm 1}) .
\label{eq:list_kernels}
\end{alignat}
\subsection{S-matrices of the $\Yf$-system}\label{app:smatrices}
The S-matrices of the basic kernels (\ref{eq:basic_kernels}) are defined as 
\begin{equation}
	-2\pi i K_a(\theta)\eqqcolon \partial _{\theta} \log S_a(\theta)
	\label{eq:def_smat_app}
\end{equation}
and are thus given by
\begin{equation}
	S_1(\theta)=i\frac{1-ie^\theta}{1+ie^\theta},\quad S_2(\theta)=\frac{2i\sinh\theta-\sqrt{2}}{2i\sinh\theta+\sqrt{2}},\quad S_3(\theta)=\cosh 2\theta.
	\label{eq:basic_smat}
\end{equation}
For the combinations of $\Yf$-system kernels (\ref{eq:list_kernels}), the S-matrices can be determined by noting that by the definition of the S-matrices, the S-matrix of a general combination of kernels $\mathcal{K} = c_{a_1} K_{a_1} + c_{a_2} K_{a_2}$ reads $\mathcal{S} = S_{a_1}^{c_{a_1}}\cdot S_{a_2}^{c_{a_2}}$.
With these conventions, residue contributions due to singularities of the $\Yf$-system of the type $\widetilde{\Yf}_{a,s}=-1$ crossing the integration contour schematically appear as follows:
\begin{align}
	&\int\limits_{\mathbb{R}}d\theta'\,\mathcal{K}_{s,s'}^{a,a'}(\theta-\theta'+i\varphi_s-i\varphi_{s'})\log\left(1+\widetilde{\Yf}_{a',s'}(\theta')\right)\nonumber\\
	&\quad\quad=\int\limits_{\mathbb{R}}d\theta'\left[-\partial_{\theta'}\left(\frac{\log\mathcal{S}_{s,s'}^{a,a'}(\theta-\theta'+i\varphi_s-i\varphi_{s'})}{-2\pi i}\right)\right]\log\left(1+\widetilde{\Yf}_{a',s'}(\theta')\right) \nonumber\\
	&\quad\quad = -\int\limits_{\mathbb{R}}d\theta'\frac{\log\mathcal{S}_{s,s'}^{a,a'}(\theta-\theta'+i\varphi_s-i\varphi_{s'})}{2\pi i}\frac{\partial_{\theta'}\widetilde{\Yf}_{a',s'}(\theta')}{1+\widetilde{\Yf}_{a',s'}(\theta')}.
	\label{eq:res_cont}
\end{align}
Hence, singularities crossing into the positive half-plane appear with a plus sign $\sim + \log\mathcal{S}_{s,s'}^{a,a'}$, while those crossing into the negative half-plane appear with a minus sign $\sim - \log\mathcal{S}_{s,s'}^{a,a'}$, as in eq.\ (\ref{eq:tba_ac}).
For crossing singularities of the type $\widetilde{\Yf}_{a,s}=\infty$, the signs are changed.
\section{Derivation of the BAE solution}
\label{sec:app_bae}
In this appendix, we derive the solution of the Bethe ansatz equations for the most general pattern of crossing singularities.
We have emphasized before that $-\frac{\pi}{4}\leq \mathrm{Im}\,\tilde{\theta}_{a,s,i}\leq \frac{\pi}{4}$ holds for the location of all crossing singularities.
This is important because, as explained in section \ref{sec:mrl_tba}, in the multi-Regge limit the residue contributions of kernel singularities referred to in eq.\ (\ref{eq:y_large_imag}) are negligible in the fundamental strip $-\frac{\pi}{4}\leq\mathrm{Im}\,\theta\leq\frac{\pi}{4}$.
Therefore, the endpoint conditions of crossing singularities in the fundamental strip (\ref{eq:rev_endp_cond}) are only coupled through the S-matrices and the Bethe ansatz equations corresponding to the most general crossing pattern read
\begin{equation}
	-1=\widetilde{\Yf}'_{a,s}(\tilde{\theta}_{a,s,i})=e^{-|m_{a,s}|'\cosh\tilde\theta_{a,s,i}+C'_{a,s}}\prod\limits_{a',s'}\prod\limits_{j=1}^{n_{a',s'}} \mathcal{S}^{a,a'}_{s,s'}\left(\tilde{\theta}_{a,s,i}-\tilde{\theta}_{a',s',j}+i\varphi_s-i\varphi_{s'}\right)^{\mathrm{sign}(\mathrm{Im}\,\tilde{\theta}_{a',s',j})},
	\label{eq:endp_cond}
\end{equation}
where the prime on the $\widetilde{\Yf}$-functions and the TBA parameters indicates quantities at the endpoint of the continuation, as before.
There is one such equation for each crossing singularity, i.e.\ a total of $\sum_{a',s'}n_{a',s'}$ Bethe ansatz equations.
In eq.\ (\ref{eq:endp_cond}), $a'\in\{1,2,3\}$ and $s'\in\{s-1,s,s+1\}$ due to the structure of the $\Yf$-system (\ref{eq:def_tba}).
Recall that in the kinematic regions under consideration, the driving term in eq.\ (\ref{eq:endp_cond}) vanishes since $|m_s|'\rightarrow \infty$ in the multi-Regge limit.
Thus, in order to get a finite expression on the right-hand side of the equations in the multi-Regge limit, the S-matrix factor on the right-hand side of eq.\ (\ref{eq:endp_cond}) has to diverge.
Accordingly, we can determine the endpoints of the crossing singularities by ensuring that the S-matrix factor diverges for all Bethe ansatz equations (\ref{eq:endp_cond}).
As in the main text, we parametrize the most general crossing pattern of singularities by the number $n_{a,s}^{(+/-)}$ of crossing singularities with endpoints in the positive (negative) half-plane and label the locations of the endpoints of the crossing singularities as $\tilde{\theta}^{(+/-)}_{a,s,i}$, $i=1,\dots, n_{a,s}^{(+/-)}$, so that
\begin{equation}
	-\frac{\pi}{4}\leq\mathrm{Im}\,\tilde{\theta}^{(-)}_{a,s,i}\leq 0,\quad 0\leq\mathrm{Im}\,\tilde{\theta}^{(+)}_{a,s,i}\leq\frac{\pi}{4}.
	\label{eq:range_tilde}
\end{equation}
We begin by studying the endpoint conditions for crossing singularities of the type $\widetilde{\Yf}'_{2,s}=-1$.
Note that such singularities can never have an endpoint at $\pm i\frac{\pi}{4}$,
\begin{equation}
	\tilde{\theta}^{(+/-)}_{2,s,i}\stackrel{!}{\neq}\pm i \frac{\pi}{4},
	\label{eq:y2_neq_pi4}
\end{equation}
as those points correspond to the cross ratios $u'_{2/3, s}$ (see eqs.\ (\ref{eq:rel_aux_crs_mrl}) and (\ref{eq:crs_ac})).
Due to our choice of continuation paths (\ref{eq:crs_signs}), these cross ratios are going to zero at the endpoint of the continuation, which requires the corresponding $\widetilde{\Yf}_{a,s}$-functions to be infinitesimal, as well.
A crossing singularity ending at $\pm i\frac{\pi}{4}$, however, would lead to the corresponding $\widetilde{\Yf}_{a,s}$-function being equal to $-1$ at that point.
Therefore, the endpoints $\pm i\frac{\pi}{4}$ for crossing singularities of the type $\widetilde{\Yf}'_{2,s}=-1$ are excluded by the relevant paths of continuation in the multi-Regge limit.
With this in mind, let us now study the endpoint condition for a crossing singularity of the type $\widetilde{\Yf}'_{2,s}=-1$ in the positive half-plane, for which the most general endpoint condition has the form
\begin{align}
	-1=&\widetilde{\Yf}'_{2,s}(\tilde{\theta}^{(+)}_{2,s,i})\label{eq:gen_y2_bae}\\
	=&e^{-\sqrt{2}|m_{2,s}|'\cosh\tilde{\theta}^{(+)}_{2,s,i}}\cdot\underbrace{\prod\limits_{s'}\frac{\prod\limits_{i_1=1}^{n^{(+)}_{2,s'}}\mathcal{S}_{s,s'}^{2,2}\left(\tilde{\theta}^{(+)}_{2,s,i}-\tilde{\theta}^{(+)}_{2,s',i_1}+i\varphi_s-i\varphi_{s'}\right)}{\prod\limits_{i_2=1}^{n^{(-)}_{2,s'}}\mathcal{S}_{s,s'}^{2,2}\left(\tilde{\theta}^{(+)}_{2,s,i}-\tilde{\theta}^{(-)}_{2,s',i_2}+i\varphi_s-i\varphi_{s'}\right)}}_{\Circled{1}}\cdot\prod\limits_{a'\in\{1,3\}}\underbrace{\frac{\prod\limits_{i_3=1}^{n_{a',s}^{(+)}}\mathcal{S}_{s,s}^{2,a'}\left(\tilde{\theta}^{(+)}_{2,s,i}-\tilde{\theta}^{(+)}_{a',s,i_3}\right)}{\prod\limits_{i_4=1}^{n^{(-)}_{a',s}}\mathcal{S}_{s,s}^{2,a'}\left(\tilde{\theta}^{(+)}_{2,s,i}-\tilde{\theta}^{(-)}_{a',s,i_4}\right)}}_{\Circled{2}} \nonumber\\
	&\cdot\underbrace{\frac{\prod\limits_{i_5=1}^{n^{(+)}_{a',s-1}}\mathcal{S}_{s,s-1}^{2,a'}\left(\tilde{\theta}^{(+)}_{2,s,i}-\tilde{\theta}^{(+)}_{a',s-1,i_5}+i\varphi_s-i\varphi_{s-1}\right)}{\prod\limits_{i_6=1}^{n^{(-)}_{a',s-1}}\mathcal{S}_{s,s-1}^{2,a'}\left(\tilde{\theta}^{(+)}_{2,s,i}-\tilde{\theta}^{(-)}_{a',s-1,i_6}+i\varphi_s-i\varphi_{s-1}\right)}}_{\Circled{3}}\cdot\underbrace{\frac{\prod\limits_{i_7=1}^{n_{a',s+1}^{(+)}}\mathcal{S}_{s,s+1}^{2,a'}\left(\tilde{\theta}^{(+)}_{2,s,i}-\tilde{\theta}^{(+)}_{a',s+1,i_7}+i\varphi_s-i\varphi_{s+1}\right)}{\prod\limits_{i_8=1}^{n^{(-)}_{a',s+1}}\mathcal{S}_{s,s+1}^{2,a'}\left(\tilde{\theta}^{(+)}_{2,s,i}-\tilde{\theta}^{(-)}_{a',s+1,i_8}+i\varphi_s-i\varphi_{s+1}\right)}}_{\Circled{4}}.\nonumber
\end{align}
We have grouped the various contributions of the S-matrix factor in eq.\ (\ref{eq:gen_y2_bae}) in four terms.
The possible endpoints of $\tilde{\theta}_{2,s,i}$ can now be determined by examining the locations of the poles of each of the four terms.
To that end, we use the expressions for the S-matrices spelled out in appendix \ref{app:smatrices}.
Note that, while all S-matrices appearing in eq.\ (\ref{eq:gen_y2_bae}) have infinitely many poles and/or zeros, the range of the argument is restricted by eq.\ (\ref{eq:range_tilde}).
Indeed, the difference between the imaginary parts of two crossing singularities in the same half-plane (in different half-planes) cannot exceed $\pm i\frac{\pi}{4}$ ($\pm i\frac{\pi}{2}$), and we only consider poles and zeros of the S-matrices in this region in the following.
We obtain the following results: 
\begin{enumerate}[label=\protect\Circled{\arabic*}]
	\item For the S-matrices $\mathcal{S}_{s, s'}^{2, 2}$, which explicitly read \begin{align*}
			&\mathcal{S}_{s, s-1}^{2,2}(x+i\varphi_s-i\varphi_{s-1})=S_2\left(x-i\frac{\pi}{4}\right)=\frac{2i\sinh\left(x-i\frac{\pi}{4}\right)-\sqrt{2}}{2i\sinh\left(x-i\frac{\pi}{4}\right)+\sqrt{2}},\\
			&\mathcal{S}_{s,s}^{2,2}(x) = S_1^{-2}(x)=-\left(\frac{1+i\,e^x}{1-i\,e^x}\right)^2,\\
			&\mathcal{S}_{s,s+1}^{2,2}(x+i\varphi_s-i\varphi_{s+1})=S_2\left(x+i\frac{\pi}{4}\right)=\frac{2i\sinh\left(x+i\frac{\pi}{4}\right)-\sqrt{2}}{2i\sinh\left(x+i\frac{\pi}{4}\right)+\sqrt{2}},
		\end{align*} the poles and zeros are located at $x=\pm i \frac{\pi}{2}$, where we have used the results of appendix \ref{app:smatrices} and eq.\ (\ref{eq:mrl_aux}). Due to the range of the endpoints (\ref{eq:range_tilde}), the only way to reach this point would be that $\tilde{\theta}^{(+)}_{2,s,i}-\tilde{\theta}^{(-)}_{2,s,j}=i\frac{\pi}{2}$ holds for some $j$, which would entail $\tilde{\theta}^{(+)}_{2,s,i}=-\tilde{\theta}^{(-)}_{2,s,j}=i\frac{\pi}{4}$. This configuration, however, is not allowed as explained around eq.\ (\ref{eq:y2_neq_pi4}). Therefore, this term does not lead to a possible endpoint for $\tilde{\theta}^{(+)}_{2,s,i}$.
	\item In this term (and the following), we have grouped the cases $a'=1$ and $a'=3$ for $s'=s-1$ as the S-matrices are identical and read \begin{align*} \mathcal{S}_{s,s}^{2,1/3}(x) = S_2^{-1}(x)=\frac{2i\sinh x+\sqrt{2}}{2i\sinh x-\sqrt{2}}. \end{align*}
		The relevant locations of the poles and zeros of this term are a pole of the S-matrices appearing in the numerator at $\tilde{\theta}^{(+)}_{2,s,i}-\tilde{\theta}_{1/3, s, j}^{(+)}=-i\frac{\pi}{4}$ and a zero of the S-matrices in the denominator at $\tilde{\theta}^{(+)}_{2,s,i}-\tilde{\theta}^{(-)}_{1/3, s, j}=i\frac{\pi}{4}$. These constraints would require a solution of $\widetilde{\Yf}_{2,s,i}=-1$ to be located a distance of $i\frac{\pi}{4}$ above (below) a solution of $\widetilde{\Yf}_{1/3,s,j}=-1$ which has crossed into the negative (positive) half-plane. However, by the analysis of section \ref{sec:related_singularities} we know that $\widetilde{\Yf}_{2,s}$ equals infinity at those points rather than $-1$. Therefore, the constraints arising from this part of the S-matrix factor cannot be satisfied and do not lead to a possible endpoint of $\tilde{\theta}^{(+)}_{2,s,i}$.
	\item  The S-matrices appearing in this term, \begin{align*} \mathcal{S}_{s,s-1}^{2,1/3}(x+i\varphi_s-i\varphi_{s-1})=S_1\left(x-i\frac{\pi}{4}\right)=i\frac{1-i\,e^{x-i\frac{\pi}{4}}}{1+i\,e^{x-i\frac{\pi}{4}}}, \end{align*} have a zero at $\tilde{\theta}^{(+)}_{2,s,i}-\tilde{\theta}^{(+/-)}_{1/3,s-1,j}=-i\frac{\pi}{4}$. Thus, a zero in the denominator could, in principle, lead to a pole in the overall S-matrix factor. However, the difference $\tilde{\theta}^{(+)}_{2,s,i}-\tilde{\theta}^{(-)}_{1/3,s-1,j}$ appearing in the denominator has a positive imaginary part by eq.\ (\ref{eq:range_tilde}) and thus the zero of the S-matrix cannot be reached by that argument. The S-matrices appearing in the numerator cannot lead to a pole of the overall S-matrix factor, as the S-matrices of this term only have zeros. Hence, this term also does not lead to a possible endpoint for $\tilde{\theta}^{(+)}_{2,s,i}$.
	\item The S-matrices appearing in this term, \begin{align*}\mathcal{S}_{s,s+1}^{2,1/3}(x+i\varphi_s-i\varphi_{s+1})=S_1\left(x+i\frac{\pi}{4}\right)=i\frac{1-i\,e^{x+i\frac{\pi}{4}}}{1+i\,e^{x+i\frac{\pi}{4}}} ,\end{align*} only have poles in the relevant range of $x$, hence only the terms appearing in the numerator are relevant for a pole of the overall S-matrix factor. This pole is located at $\tilde{\theta}^{(+)}_{2,s,i}-\tilde{\theta}^{(+)}_{1/3,s+1,j}=i\frac{\pi}{4}$ for some $j$. However, due to the range of the variables (\ref{eq:range_tilde}) this condition would enforce $\tilde{\theta}^{(+)}_{2,s,i}=i\frac{\pi}{4}$, which is not possible as explained for term $\Circled{1}$ above. Therefore, this term does not lead to a possible endpoint for $\tilde{\theta}^{(+)}_{2,s,i}$. 
\end{enumerate}
While we have only provided the details for singularities of the type $\widetilde{\Yf}'_{2,s}=-1$ crossing into the positive half-plane, the results also hold equivalently for singularities crossing into the negative half-plane.
Hence, in the multi-Regge limit the endpoint condition for crossing singularities of the type $\widetilde{\Yf}'_{2,s}=-1$ cannot be satisfied, as there is no allowed endpoint of such a crossing singularity which would approach a pole of the S-matrix factor and could compensate the small driving term in eq.\ (\ref{eq:gen_y2_bae}).
Therefore, these types of crossing singularities can be neglected in the following.\par
We now turn to the cases $a=1, 3$, but refrain from spelling out all details as the analysis is very similar to that of eq.\ (\ref{eq:gen_y2_bae}).
The key difference, however, is that in these cases there are locations of the poles of the S-matrix factor which can be reached by the endpoints of the crossing singularities.
For example, from the analysis of the most general endpoint condition of a singularity of the type $\widetilde{\Yf}_{1,s}(\tilde{\theta}^{(+)}_{1,s,i})=-1$ in the positive half-plane, we obtain the following possible configurations which lead to a pole in the S-matrix factor:
\begin{itemize}
	\item $\tilde{\theta}^{(+)}_{1,s,i}-\tilde{\theta}^{(-)}_{3, s-1, j} = 0$ (which, together with eq.\ (\ref{eq:range_tilde}), implies $\tilde{\theta}^{(+)}_{1,s,i}=\tilde{\theta}^{(-)}_{3, s-1, j} = 0$),
	\item $\tilde{\theta}^{(+)}_{1,s,i}-\tilde{\theta}^{(-)}_{1, s, j} = i\frac{\pi}{2}$ (which implies $\tilde{\theta}^{(+)}_{1,s,i}=-\tilde{\theta}^{(-)}_{1, s, j} = i\frac{\pi}{4}$),\footnote{Strictly speaking, this condition implies $\tilde{\theta}^{(+)}_{1,s,i}=x+i\frac{\pi}{4}$, $\tilde{\theta}^{(-)}_{1, s, j} =x-i\frac{\pi}{4}$, for some $x\in\mathbb{R}$, only. However, the most general BAE (\ref{eq:endp_cond}) can be schematically rewritten as
	\begin{equation}
		k\cdot i\pi = -|m_{a,s}|'\cosh\,\tilde{\theta}_{a,s,i}+C'_{a,s}+\log\mathcal{S}(\{\tilde{\theta}_{a', s', j}\}),
	\label{eq:cond_im_zero}
	\end{equation}
	where $k\in\mathbb{Z}$ and where we have collected all S-matrices into a single factor.
	These are two equations for each crossing singularity -- one for the real part and one for the imaginary part of eq.\ (\ref{eq:cond_im_zero}).
	As explained in the main text, since $\mathrm{Re}\,|m_{a,s}|'\rightarrow \infty$ in the multi-Regge limit, the S-matrix factor has to diverge.
	However, as the poles of the S-matrices appearing in eq.\ (\ref{eq:cond_im_zero}) are all simple poles, the singularities appearing in that equation are logarithmic singularities, at which the imaginary part is discontinuous, but does not diverge.
	The condition, that the imaginary part of eq.\ (\ref{eq:cond_im_zero}) has to remain finite in the multi-Regge limit then implies that $\mathrm{Re}\,\tilde{\theta}_{a,s,i} \stackrel{!}{=} 0$, which leads to the implication stated in the main text.
	}
	\item $\tilde{\theta}^{(+)}_{1,s,i}-\tilde{\theta}^{(-)}_{3, s, j} = i\frac{\pi}{2}$ (which implies $\tilde{\theta}^{(+)}_{1,s,i}=-\tilde{\theta}^{(-)}_{3, s, j} = i\frac{\pi}{4}$) or
	\item $\tilde{\theta}^{(+)}_{1,s,i}-\tilde{\theta}^{(+)}_{3, s+1, j} = 0$.
\end{itemize}
Importantly, the endpoint condition does not uniquely specify the endpoint of the crossing singularity $\tilde{\theta}^{(+)}_{1,s,i}$.
However, there are additional constraints from the multi-Regge behavior of the cross ratios.
By eq.\ (\ref{eq:crs_ac}), the cross ratios $u_{2/3, s}'$ at the endpoint of the continuation are related to the $\widetilde{\Yf}_{2,s}$-functions at $\theta=\pm i\frac{\pi}{4}$ and are therefore also subject to S-matrix contributions from the crossing singularities.
For the most general pattern of crossing singularities, these contributions read
\begin{align}
	u'_{2/3,s}\approx &\frac{u'_{2/3,s}}{1-u'_{2/3,s}}=\widetilde{\Yf}'_{2,s}\left(\pm i\frac{\pi}{4}\right)\nonumber\\
	=&e^{-\sqrt{2}|m_s|'}\cdot\prod\limits_{a'\in\{1,3\}}\frac{\prod\limits_{i_1=1}^{n^{(+)}_{a', s-1}}\mathcal{S}_{s,s-1}^{2, a'}\left(\pm i\frac{\pi}{4}-\tilde{\theta}^{(+)}_{a',s-1,i_1}+i\varphi_s-i\varphi_{s-1}\right)}{\prod\limits_{i_2=1}^{n^{(-)}_{a', s-1}}\mathcal{S}_{s,s-1}^{2, a'}\left(\pm i\frac{\pi}{4}-\tilde{\theta}^{(-)}_{a',s-1,i_2}+i\varphi_s-i\varphi_{s-1}\right)}\nonumber\\
	&\cdot\frac{\prod\limits_{i_3=1}^{n^{(+)}_{a', s}}\mathcal{S}_{s,s}^{2, a'}\left(\pm i\frac{\pi}{4}-\tilde{\theta}^{(+)}_{a',s,i_3}\right)}{\prod\limits_{i_4=1}^{n^{(-)}_{a', s}}\mathcal{S}_{s,s}^{2, a'}\left(\pm i\frac{\pi}{4}-\tilde{\theta}^{(-)}_{a',s,i_4}\right)}\cdot\frac{\prod\limits_{i_5=1}^{n^{(+)}_{a', s+1}}\mathcal{S}_{s,s+1}^{2, a'}\left(\pm i\frac{\pi}{4}-\tilde{\theta}^{(+)}_{a',s+1,i_5}+i\varphi_s-i\varphi_{s+1}\right)}{\prod\limits_{i_6=1}^{n^{(-)}_{a', s+1}}\mathcal{S}_{s,s+1}^{2, a'}\left(\pm i\frac{\pi}{4}-\tilde{\theta}^{(-)}_{a',s+1,i_6}+i\varphi_s-i\varphi_{s+1}\right)}.
	\label{eq:general_y2_crs}
\end{align}
Note that in eq.\ (\ref{eq:general_y2_crs}), we have already neglected contributions from crossing singularities of the type $\widetilde{\Yf}_{2,s}=-1$ as those cannot occur, as explained around eq.\ (\ref{eq:gen_y2_bae}).
By our choice of endpoints of the analytic continuation (\ref{eq:crs_signs}), the cross ratios $u_{2/3, s}'$ have to approach zero at the same rate as the cross ratios $u_{2/3,s}$ at the starting point of the analytic continuation. 
Therefore, in the multi-Regge limit the endpoints of the crossing singularities may never end on a zero or a pole of the S-matrix factor in eq.\ (\ref{eq:general_y2_crs}), which introduces constraints on the endpoints of the crossing singularities.
An analysis similar to that around eq.\ (\ref{eq:gen_y2_bae}) shows that this constraint excludes zero as an endpoint for any crossing singularity,\footnote{In principle, eq.\ (\ref{eq:general_y2_crs}) allows pairs of crossing singularities (i.e.\ one crossing singularity in the positive and one crossing singularity in the negative half-plane) with the same index $s$ with both endpoints at zero. However, the contribution of such pairs to both the cross ratios in eq.\ (\ref{eq:general_y2_crs}) and the $A'_\mathrm{free}$-contribution in eq.\ (\ref{eq:a_free_ac}) vanishes, so that from the point of view of the remainder function such configurations would not be visible and are hence neglected in the following.} i.e.\ $\tilde{\theta}_{a,s,i}\neq 0$.
Using this constraint, only
\begin{equation}
	\tilde{\theta}^{(+/-)}_{1/3,s,i}=\pm i\frac{\pi}{4}
\end{equation}
remains as a possible endpoint for the crossing singularities,\footnote{One quick way to see this is to choose $\tilde{s}$ as the largest $s$-index for which there is a crossing singularity. Then, there are no contributions of S-matrices of the form $\mathcal{S}_{\tilde{s}, \tilde{s}+1}^{2, a'}$ in eq.\ (\ref{eq:general_y2_crs}). We then choose $\theta=+i\frac{\pi}{4}$, in which case the S-matrices $\mathcal{S}_{\tilde{s}, \tilde{s}-1}^{2, a'}$ have no poles or zeros in the range (\ref{eq:range_tilde}), so that only the S-matrices $\mathcal{S}_{\tilde{s}, \tilde{s}}^{2,a'}$ are relevant. The latter explicitly enforce the constraint that $\tilde{\theta}_{a', \tilde{s}, i}\neq 0$. Of the four possible endpoint configurations listed in the main text, the fourth is not available (since there is no crossing singularity with index $\tilde{s}+1$ by definition) and the first one is excluded by the constraints on the cross ratios, leaving only the second and third configuration which both imply the result stated in the main text. The result for channels $s<\tilde{s}$ follows similarly.} which is the result quoted in the main text.\par
Before closing this appendix, let us comment on an observation made in section \ref{sec:crossing_diamonds}, namely that the combinations of S-matrices leading to \textit{effective} crossing singularities come with an additional minus sign (see eq.\ (\ref{eq:effective_smatrix})).
This entails that, depending on the number of these effective crossing singularities, the left-hand side of the Bethe ansatz equations (\ref{eq:endp_cond}) equals $1$ instead of $-1$.
However, the arguments used in this appendix never used the explicit value on the left-hand side of the Bethe ansatz equations, but only that it is a finite value.
Since we have identified a unique configuration of endpoints for the crossing singularities, the analysis of this appendix also covers the case of the effective crossing singularities.
Also note that the configurations used in section \ref{sec:rem_fct} to extract the BFKL eigenvalues have an equal number of positive and negative crossing singularities (see discussion around eq.\ (\ref{eq:spectrum_bfkl})), in which case the additional minus sign is not present.
\section{Factorization of $A_\mathrm{per}$}
\label{app:aper}
In this appendix, we show that the $A_\mathrm{per}$-contribution to the remainder function at the endpoint of the analytic continuation factorizes, i.e.\ that there is no dependence between terms of different $s$-indices.
The contribution $A_\mathrm{per}$ for a general $n$-point amplitude can be written as a polynomial in the complex parameters $m_s$,
\begin{equation}
	A_\mathrm{per} = \sum\limits_{i,j}\mathcal{K}_{ij}m_i\bar{m}_j,
	\label{eq:def_aper}
\end{equation}
where $\mathcal{K}$ is a symmetric $(n-5)\times(n-5)$-matrix, whose entries depend on the parity of $(n-1)/2$ and whose construction is described in detail in \cite{Alday:2010vh}.
To connect $A_\mathrm{per}$ with the parameters (\ref{eq:rel_aper}), we use the definition of the parameters $m_s = |m_s|e^{i\varphi_s}$ to rewrite the contribution as
\begin{align}
	A_\mathrm{per}=& 2\sum\limits_{i<j}\mathcal{K}_{ij}\cos(\varphi_i-\varphi_j)|m_i||m_j|+\sum\limits_i \mathcal{K}_{ii}|m_i|^2 \label{eq:aper_eps_w}\\
	=&2\sum\limits_{i<j}\mathcal{K}_{ij}\left(\log\varepsilon_i\log\varepsilon_j\cos\left( (j-i)\frac{\pi}{4}\right)+\log w_i\log w_j \cos\left( (j-i)\frac{\pi}{4}\right)\right.\nonumber\\
	&\left.\quad+\log w_i\log\varepsilon_j\sin\left( (j-i)\frac{\pi}{4}\right)-\log\varepsilon_i\log w_j\sin\left( (j-i)\frac{\pi}{4}\right)\right)+\sum\limits_{i}\mathcal{K}_{ii}\left(\log^2\varepsilon_i+\log^2 w_i\right),\nonumber
\end{align}
where in the second step we have used the definition of the kinematic parameters (\ref{eq:def_mrl_param}).
Based on this expression, it is straightforward to calculate the contribution $A'_{\mathrm{per}}(\varepsilon'_s, w'_s)-A_{\mathrm{per}}(\varepsilon_s, w_s)$ to the remainder function at the endpoint of the analytic continuation.
Using the expressions (\ref{eq:rel_aper}), we obtain the result used in the main text
\begin{align}
	A'_{\mathrm{per}}&(\varepsilon'_s, w'_s)-A_{\mathrm{per}}(\varepsilon_s, w_s)=\nonumber\\
	&\quad\frac{1}{2}\log(1+\sqrt{2})\sum\limits_s\left(-\log\varepsilon_s (n_{1/3, s}^{(+)}+n_{1/3,s}^{(-)})+\log w_s(n_{1/3,s}^{(-)}-n_{1/3,s}^{(+)})\right)+\dots,
	\label{eq:delta_aper}
\end{align}
where the dots indicate phases and constants, which we cannot fix as explained above.
Hence, terms including different values of the index $s$ indeed cancel and the contribution $A_{\mathrm{per}}$ to the remainder function factorizes.
However, this cancellation of terms with different $s$-indices from eq.\ (\ref{eq:aper_eps_w}) to eq.\ (\ref{eq:delta_aper}) is not obvious.
Using the explicit form of the matrices $\mathcal{K}_{ij}$, we generated the corresponding expressions (\ref{eq:aper_eps_w}) and checked that the relation (\ref{eq:delta_aper}) holds for all cases $n\leq 30$, $n\neq 4k$.\footnote{Note that there is a small typo in the matrix $\mathcal{K}_3$ used in \cite{Alday:2010vh} for the case $n=4k+5$, namely the entry $(\mathcal{K}_3)_{1,4}$ should read $1$, not $0$ as stated there.}
Since the cancellations between terms with different $s$-indices hinges on the explicit entries of the matrices $\mathcal{K}_{ij}$ an analytic proof for all values of $n$ does not seem feasible.
\section{Kinematic parameters in subleading kinematics}
\label{app:subleading}
In section \ref{sec:consistency}, we analyze a set of contributions to the remainder function which are kinematically subleading in the multi-Regge limit.
For this analysis, the first subleading order of the kinematics parameters $\varepsilon_s'$ and $w_s'$ at the endpoints is needed, which reads:
\small
\begin{align}
\varepsilon_1'&=\varepsilon_1\left(1-\frac{2 \sqrt{2} \log w_2}{\log \varepsilon_2}+\mathcal{O}(\log^{-2}\varepsilon)\right),\nonumber\\
w_1'&=\gamma\, w_1 \left(1-\frac{4 \sqrt{2} \log(1+\sqrt{2}\,)}{\log\varepsilon_1}-\frac{2 \sqrt{2} \log w_1}{\log\varepsilon_1}+\frac{2 \sqrt{2} \log w_2}{\log\varepsilon_2}+\mathcal{O}(\log^{-2}\varepsilon)\right),\nonumber\\
\varepsilon_2'&=\frac{1}{\gamma}\,\varepsilon_2 \left(1+\frac{2 \sqrt{2} \log(1+\sqrt{2}\,)}{\log\varepsilon_1}+\frac{\sqrt{2} \log w_1}{\log\varepsilon_1}+\frac{2 \sqrt{2} \log(1+\sqrt{2}\,)}{\log\varepsilon_3}-\frac{\sqrt{2} \log w_3}{\log \varepsilon_3}+\mathcal{O}(\log^{-2}\varepsilon)\right),\nonumber\\
w_2'&=w_2 \left(1+\frac{2 \sqrt{2} \log(1+\sqrt{2}\,)}{\log\varepsilon_1}+\frac{\sqrt{2} \log w_1}{\log\varepsilon_1}-\frac{4 \sqrt{2} \log w_2}{\log\varepsilon_2}-\frac{2 \sqrt{2} \log(1+\sqrt{2}\,)}{\log\varepsilon_3}+\frac{\sqrt{2} \log w_3}{\log\varepsilon_3}+\mathcal{O}(\log^{-2}\varepsilon)\right),\nonumber\\
\varepsilon_3'&=\varepsilon_3 \left(1+\frac{2 \sqrt{2} \log w_2}{\log\varepsilon_2}+\mathcal{O}(\log^{-2}\varepsilon)\right),\nonumber\\
w_3'&=\frac{1}{\gamma}\,w_3 \left(1+\frac{2 \sqrt{2} \log w_2}{\log\varepsilon_2}+\frac{4 \sqrt{2} \log(1+\sqrt{2}\,)}{\log\varepsilon_3}-\frac{2 \sqrt{2} \log w_3}{\log\varepsilon_3}+\mathcal{O}(\log^{-2}\varepsilon)\right),\nonumber\\
\varepsilon_4'&=\sqrt{\gamma}\,\varepsilon_4 \left(1-\frac{2 \sqrt{2} \log(1+\sqrt{2})}{\log\varepsilon_3}+\frac{\sqrt{2} \log w_3}{\log\varepsilon_3}+\mathcal{O}(\log^{-2}\varepsilon)\right),\nonumber\\
w_4'&=\frac{1}{\sqrt{\gamma}}\,w_4 \left(1-\frac{2 \sqrt{2} \log (1+\sqrt{2})}{\log\varepsilon_3}+\frac{\sqrt{2} \log w_3}{\log\varepsilon_3}+\mathcal{O}(\log^{-2}\varepsilon)\right).
\end{align}
\normalsize

\bibliographystyle{JHEP}
\bibliography{arxiv_final}
\end{document}